

%
%
%
\def\unredoffs{} \def\redoffs{\voffset=-.31truein\hoffset=-.48truein}
\def\speclscape{}
%
%
%
%
%

\newbox\leftpage \newdimen\fullhsize \newdimen\hstitle \newdimen\hsbody
\tolerance=1000\hfuzz=2pt
\catcode`\@=11 
\ifx\hyperdef\UNd@FiNeD\def\hyperdef#1#2#3#4{#4}\def\hyperref#1#2#3#4{#4}\fi
\def\bigans{b }
\def\answ{b }
%
\ifx\answ\bigans\message{(This will come out unreduced.}
\magnification=1200\unredoffs\baselineskip=16pt plus 2pt minus 1pt
\hsbody=\hsize \hstitle=\hsize 
\else\message{(This will be reduced.} \let\l@r=L
\magnification=1000\baselineskip=16pt plus 2pt minus 1pt \vsize=7truein
\redoffs \hstitle=8truein\hsbody=4.75truein\fullhsize=10truein\hsize=\hsbody
\output={\ifnum\pageno=0 
  \shipout\vbox{\speclscape{\hsize\fullhsize\makeheadline}
    \hbox to \fullhsize{\hfill\pagebody\hfill}}\advancepageno
  \else
  \almostshipout{\leftline{\vbox{\pagebody\makefootline}}}\advancepageno
  \fi}
\def\almostshipout#1{\if L\l@r \count1=1 \message{[\the\count0.\the\count1]}
      \global\setbox\leftpage=#1 \global\let\l@r=R
 \else \count1=2
  \shipout\vbox{\speclscape{\hsize\fullhsize\makeheadline}
      \hbox to\fullhsize{\box\leftpage\hfil#1}}  \global\let\l@r=L\fi}
\fi
%
\newcount\yearltd\yearltd=\year\advance\yearltd by -2000

\def\Title#1#2{\nopagenumbers\abstractfont\hsize=\hstitle\rightline{#1}%
\vskip 1in\centerline{\titlefont #2}\abstractfont\vskip .5in\pageno=0}
\def\Date#1{\vfill\leftline{#1}\tenpoint\supereject\global\hsize=\hsbody%
\footline={\hss\tenrm\hyperdef\hypernoname{page}\folio\folio\hss}}%
%

\def\draftmode{\message{ DRAFTMODE }\def\draftdate{{\rm preliminary draft:
\number\month/\number\day/\number\yearltd\ \ \hourmin}}%
\headline={\hfil\draftdate}\writelabels\baselineskip=20pt plus 2pt minus 2pt
 {\count255=\time\divide\count255 by 60 \xdef\hourmin{\number\count255}
  \multiply\count255 by-60\advance\count255 by\time
  \xdef\hourmin{\hourmin:\ifnum\count255<10 0\fi\the\count255}}}
\def\nolabels{\def\wrlabeL##1{}\def\eqlabeL##1{}\def\reflabeL##1{}}
\def\writelabels{\def\wrlabeL##1{\leavevmode\vadjust{\rlap{\smash%
{\line{{\escapechar=` \hfill\rlap{\sevenrm\hskip.03in\string##1}}}}}}}%
\def\eqlabeL##1{{\escapechar-1\rlap{\sevenrm\hskip.05in\string##1}}}%
\def\reflabeL##1{\noexpand\llap{\noexpand\sevenrm\string\string\string##1}}}
\nolabels
%
\global\newcount\secno \global\secno=0
\global\newcount\meqno \global\meqno=1
\def\s@csym{}
\def\newsec#1{\global\advance\secno by1%
{\toks0{#1}\message{(\the\secno. \the\toks0)}}%
\global\subsecno=0\eqnres@t\let\s@csym\secsym\xdef\secn@m{\the\secno}\noindent
{\bf\hyperdef\hypernoname{section}{\the\secno}{\the\secno.} #1}%
\writetoca{{\string\hyperref{}{section}{\the\secno}{\the\secno.}} {#1}}%
\par\nobreak\medskip\nobreak}
\def\eqnres@t{\xdef\secsym{\the\secno.}\global\meqno=1\bigbreak\bigskip}
\def\sequentialequations{\def\eqnres@t{\bigbreak}}\xdef\secsym{}
\global\newcount\subsecno \global\subsecno=0
\def\subsec#1{\global\advance\subsecno by1%
{\toks0{#1}\message{(\s@csym\the\subsecno. \the\toks0)}}%
\ifnum\lastpenalty>9000\else\bigbreak\fi
\noindent{\it\hyperdef\hypernoname{subsection}{\secn@m.\the\subsecno}%
{\secn@m.\the\subsecno.} #1}\writetoca{\string\quad
{\string\hyperref{}{subsection}{\secn@m.\the\subsecno}{\secn@m.\the\subsecno.}}
{#1}}\par\nobreak\medskip\nobreak}
\def\appendix#1#2{\global\meqno=1\global\subsecno=0\xdef\secsym{\hbox{#1.}}%
\bigbreak\bigskip\noindent{\bf Appendix \hyperdef\hypernoname{appendix}{#1}%
{#1.} #2}{\toks0{(#1. #2)}\message{\the\toks0}}%
\xdef\s@csym{#1.}\xdef\secn@m{#1}%
\writetoca{\string\hyperref{}{appendix}{#1}{Appendix {#1.}} {#2}}%
\par\nobreak\medskip\nobreak}
%
%
\def\checkm@de#1#2{\ifmmode{\def\f@rst##1{##1}\hyperdef\hypernoname{equation}%
{#1}{#2}}\else\hyperref{}{equation}{#1}{#2}\fi}
\def\eqnn#1{\DefWarn#1\xdef #1{(\noexpand\relax\noexpand\checkm@de%
{\s@csym\the\meqno}{\secsym\the\meqno})}%
\wrlabeL#1\writedef{#1\leftbracket#1}\global\advance\meqno by1}
\def\f@rst#1{\c@t#1a\em@ark}\def\c@t#1#2\em@ark{#1}
\def\eqna#1{\DefWarn#1\wrlabeL{#1$\{\}$}%
\xdef #1##1{(\noexpand\relax\noexpand\checkm@de%
{\s@csym\the\meqno\noexpand\f@rst{##1}}{\hbox{$\secsym\the\meqno##1$}})}
\writedef{#1\numbersign1\leftbracket#1{\numbersign1}}\global\advance\meqno by1}
\def\eqn#1#2{\DefWarn#1%
\xdef #1{(\noexpand\hyperref{}{equation}{\s@csym\the\meqno}%
{\secsym\the\meqno})}$$#2\eqno(\hyperdef\hypernoname{equation}%
{\s@csym\the\meqno}{\secsym\the\meqno})\eqlabeL#1$$%
\writedef{#1\leftbracket#1}\global\advance\meqno by1}
\def\xeqn{\expandafter\xe@n}\def\xe@n(#1){#1}
\def\xeqna#1{\expandafter\xe@n#1}
\def\eqns#1{(\e@ns #1{\hbox{}})}
\def\e@ns#1{\ifx\UNd@FiNeD#1\message{eqnlabel \string#1 is undefined.}%
\xdef#1{(?.?)}\fi{\let\hyperref=\relax\xdef\next{#1}}%
\ifx\next\em@rk\def\next{}\else%
\ifx\next#1\xeqn#1\else\def\n@xt{#1}\ifx\n@xt\next#1\else\xeqna#1\fi
\fi\let\next=\e@ns\fi\next}

\def\DefWarn#1{\ifx\UNd@FiNeD#1\else
\immediate\write16{*** WARNING: the label \string#1 is already defined ***}\fi}
%
\newskip\footskip\footskip14pt plus 1pt minus 1pt 
\def\footnotefont{\ninepoint}\def\f@t#1{\footnotefont #1\@foot}
\def\f@@t{\baselineskip\footskip\bgroup\footnotefont\aftergroup\@foot\let\next}
\setbox\strutbox=\hbox{\vrule height9.5pt depth4.5pt width0pt}
\global\newcount\ftno \global\ftno=0
\def\foot{\global\advance\ftno by1\def\foot@rg{\hyperref{}{footnote}%
{\the\ftno}{\the\ftno}\xdef\foot@rg{\noexpand\hyperdef\noexpand\hypernoname%
{footnote}{\the\ftno}{\the\ftno}}}\footnote{$^{\foot@rg}$}}
%
\newwrite\ftfile
\def\footend{\def\foot{\global\advance\ftno by1\chardef\wfile=\ftfile
\hyperref{}{footnote}{\the\ftno}{$^{\the\ftno}$}%
\ifnum\ftno=1\immediate\openout\ftfile=\jobname.fts\fi%
\immediate\write\ftfile{\noexpand\smallskip%
\noexpand\item{\noexpand\hyperdef\noexpand\hypernoname{footnote}
{\the\ftno}{f\the\ftno}:\ }\pctsign}\findarg}%
\def\footatend{\vfill\eject\immediate\closeout\ftfile{\parindent=20pt
\centerline{\bf Footnotes}\nobreak\bigskip\input \jobname.fts }}}
\def\footatend{}
%
%
\global\newcount\refno \global\refno=1
\newwrite\rfile
\def\ref{[\hyperref{}{reference}{\the\refno}{\the\refno}]\nref}
\def\nref#1{\DefWarn#1%
\xdef#1{[\noexpand\hyperref{}{reference}{\the\refno}{\the\refno}]}%
\writedef{#1\leftbracket#1}%
\ifnum\refno=1\immediate\openout\rfile=\jobname.refs\fi
\chardef\wfile=\rfile\immediate\write\rfile{\noexpand\item{[\noexpand\hyperdef%
\noexpand\hypernoname{reference}{\the\refno}{\the\refno}]\ }%
\reflabeL{#1\hskip.31in}\pctsign}\global\advance\refno by1\findarg}
\def\findarg#1#{\begingroup\obeylines\newlinechar=`\^^M\pass@rg}
{\obeylines\gdef\pass@rg#1{\writ@line\relax #1^^M\hbox{}^^M}%
\gdef\writ@line#1^^M{\expandafter\toks0\expandafter{\striprel@x #1}%
\edef\next{\the\toks0}\ifx\next\em@rk\let\next=\endgroup\else\ifx\next\empty%
\else\immediate\write\wfile{\the\toks0}\fi\let\next=\writ@line\fi\next\relax}}
\def\striprel@x#1{} \def\em@rk{\hbox{}}
\def\lref{\begingroup\obeylines\lr@f}
\def\lr@f#1#2{\DefWarn#1\gdef#1{\let#1=\UNd@FiNeD\ref#1{#2}}\endgroup\unskip}

\def\addref#1{\immediate\write\rfile{\noexpand\item{}#1}} 
\def\listrefs{\footatend\vfill\supereject\immediate\closeout\rfile\writestoppt
\baselineskip=\footskip\centerline{{\bf References}}\bigskip{\parindent=20pt%
\frenchspacing\escapechar=` \input \jobname.refs\vfill\eject}\nonfrenchspacing}
\def\startrefs#1{\immediate\openout\rfile=\jobname.refs\refno=#1}
\def\xref{\expandafter\xr@f}\def\xr@f[#1]{#1}
\def\refs#1{\count255=1[\r@fs #1{\hbox{}}]}
\def\r@fs#1{\ifx\UNd@FiNeD#1\message{reflabel \string#1 is undefined.}%
\nref#1{need to supply reference \string#1.}\fi%
\vphantom{\hphantom{#1}}{\let\hyperref=\relax\xdef\next{#1}}%
\ifx\next\em@rk\def\next{}%
\else\ifx\next#1\ifodd\count255\relax\xref#1\count255=0\fi%
\else#1\count255=1\fi\let\next=\r@fs\fi\next}
%

%
\newwrite\ffile\global\newcount\figno \global\figno=1
\def\fig{fig.~\hyperref{}{figure}{\the\figno}{\the\figno}\nfig}
\def\nfig#1{\DefWarn#1%
\xdef#1{fig.~\noexpand\hyperref{}{figure}{\the\figno}{\the\figno}}%
\writedef{#1\leftbracket fig.\noexpand~\xfig#1}%
\ifnum\figno=1\immediate\openout\ffile=\jobname.figs\fi\chardef\wfile=\ffile%
{\let\hyperref=\relax
\immediate\write\ffile{\noexpand\medskip\noexpand\item{Fig.\ %
\noexpand\hyperdef\noexpand\hypernoname{figure}{\the\figno}{\the\figno}. }
\reflabeL{#1\hskip.55in}\pctsign}}\global\advance\figno by1\findarg}
\def\listfigs{\vfill\eject\immediate\closeout\ffile{\parindent40pt
\baselineskip14pt\centerline{{\bf Figure Captions}}\nobreak\medskip
\escapechar=` \input \jobname.figs\vfill\eject}}
\def\xfig{\expandafter\xf@g}\def\xf@g fig.\penalty\@M\ {}
\def\figs#1{figs.~\f@gs #1{\hbox{}}}
\def\f@gs#1{{\let\hyperref=\relax\xdef\next{#1}}\ifx\next\em@rk\def\next{}\else
\ifx\next#1\xfig #1\else#1\fi\let\next=\f@gs\fi\next}
\def\figin{\epsfcheck\figin}\def\figins{\epsfcheck\figins}
\def\epsfcheck{\ifx\epsfbox\UNd@FiNeD
\message{(NO epsf.tex, FIGURES WILL BE IGNORED)}
\gdef\figin##1{\vskip2in}\gdef\figins##1{\hskip.5in}
\else\message{(FIGURES WILL BE INCLUDED)}%
\gdef\figin##1{##1}\gdef\figins##1{##1}\fi}
\def\DefWarn#1{}
\def\figinsert{\goodbreak\midinsert}
\def\ifig#1#2#3{\DefWarn#1\xdef#1{fig.~\noexpand\hyperref{}{figure}%
{\the\figno}{\the\figno}}\writedef{#1\leftbracket fig.\noexpand~\xfig#1}%
\figinsert\figin{\centerline{#3}}\medskip\centerline{\vbox{\baselineskip12pt
\advance\hsize by -1truein\noindent\wrlabeL{#1=#1}\footnotefont%
{\bf Fig.~\hyperdef\hypernoname{figure}{\the\figno}{\the\figno}:} #2}}
\bigskip\endinsert\global\advance\figno by1}
\newwrite\lfile
{\escapechar-1\xdef\pctsign{\string\%}\xdef\leftbracket{\string\{}
\xdef\rightbracket{\string\}}\xdef\numbersign{\string\#}}
\def\writedefs{\immediate\openout\lfile=\jobname.defs \def\writedef##1{%
{\let\hyperref=\relax\let\hyperdef=\relax\let\hypernoname=\relax
 \immediate\write\lfile{\string\def\string##1\rightbracket}}}}%
\def\writestop{\def\writestoppt{\immediate\write\lfile{\string\pageno
 \the\pageno\string\startrefs\leftbracket\the\refno\rightbracket
 \string\def\string\secsym\leftbracket\secsym\rightbracket
 \string\secno\the\secno\string\meqno\the\meqno}\immediate\closeout\lfile}}
\def\writestoppt{}\def\writedef#1{}
\def\seclab#1{\DefWarn#1%
\xdef #1{\noexpand\hyperref{}{section}{\the\secno}{\the\secno}}%
\writedef{#1\leftbracket#1}\wrlabeL{#1=#1}}
\def\subseclab#1{\DefWarn#1%
\xdef #1{\noexpand\hyperref{}{subsection}{\secn@m.\the\subsecno}%
{\secn@m.\the\subsecno}}\writedef{#1\leftbracket#1}\wrlabeL{#1=#1}}
\def\applab#1{\DefWarn#1%
\xdef #1{\noexpand\hyperref{}{appendix}{\secn@m}{\secn@m}}%
\writedef{#1\leftbracket#1}\wrlabeL{#1=#1}}
\newwrite\tfile \def\writetoca#1{}
\def\leaderfill{\leaders\hbox to 1em{\hss.\hss}\hfill}
\def\writetoc{\immediate\openout\tfile=\jobname.toc
   \def\writetoca##1{{\edef\next{\write\tfile{\noindent ##1
   \string\leaderfill {\string\hyperref{}{page}{\noexpand\number\pageno}%
                       {\noexpand\number\pageno}} \par}}\next}}}
\newread\ch@ckfile
\def\listtoc{\immediate\closeout\tfile\immediate\openin\ch@ckfile=\jobname.toc
\ifeof\ch@ckfile\message{no file \jobname.toc, no table of contents this pass}%
\else\closein\ch@ckfile\centerline{\bf Contents}\nobreak\medskip%
{\baselineskip=12pt\footnotefont\parskip=0pt\catcode`\@=11\input\jobname.toc
\catcode`\@=12\bigbreak\bigskip}\fi}
\catcode`\@=12 
%
\edef\tfontsize{\ifx\answ\bigans scaled\magstep3\else scaled\magstep4\fi}
\font\titlerm=cmr10 \tfontsize \font\titlerms=cmr7 \tfontsize
\font\titlermss=cmr5 \tfontsize \font\titlei=cmmi10 \tfontsize
\font\titleis=cmmi7 \tfontsize \font\titleiss=cmmi5 \tfontsize
\font\titlesy=cmsy10 \tfontsize \font\titlesys=cmsy7 \tfontsize
\font\titlesyss=cmsy5 \tfontsize \font\titleit=cmti10 \tfontsize
\skewchar\titlei='177 \skewchar\titleis='177 \skewchar\titleiss='177
\skewchar\titlesy='60 \skewchar\titlesys='60 \skewchar\titlesyss='60
\def\titlefont{\def\rm{\fam0\titlerm}
\textfont0=\titlerm \scriptfont0=\titlerms \scriptscriptfont0=\titlermss
\textfont1=\titlei \scriptfont1=\titleis \scriptscriptfont1=\titleiss
\textfont2=\titlesy \scriptfont2=\titlesys \scriptscriptfont2=\titlesyss
\textfont\itfam=\titleit \def\it{\fam\itfam\titleit}\rm}
 \ifx\answ\bigans\else scaled\magstep1\fi
\ifx\answ\bigans\def\abstractfont{\tenpoint}\else
\font\absit=cmti10 scaled \magstep1
\font\abssl=cmsl10 scaled \magstep1
\font\absrm=cmr10 scaled\magstep1 \font\absrms=cmr7 scaled\magstep1
\font\absrmss=cmr5 scaled\magstep1 \font\absi=cmmi10 scaled\magstep1
\font\absis=cmmi7 scaled\magstep1 \font\absiss=cmmi5 scaled\magstep1
\font\abssy=cmsy10 scaled\magstep1 \font\abssys=cmsy7 scaled\magstep1
\font\abssyss=cmsy5 scaled\magstep1 \font\absbf=cmbx10 scaled\magstep1
\skewchar\absi='177 \skewchar\absis='177 \skewchar\absiss='177
\skewchar\abssy='60 \skewchar\abssys='60 \skewchar\abssyss='60
\def\abstractfont{\def\rm{\fam0\absrm}
\textfont0=\absrm \scriptfont0=\absrms \scriptscriptfont0=\absrmss
\textfont1=\absi \scriptfont1=\absis \scriptscriptfont1=\absiss
\textfont2=\abssy \scriptfont2=\abssys \scriptscriptfont2=\abssyss
\textfont\itfam=\absit \def\it{\fam\itfam\absit}\def\footnotefont{\tenpoint}%
\textfont\slfam=\abssl \def\sl{\fam\slfam\abssl}%
\textfont\bffam=\absbf \def\bf{\fam\bffam\absbf}\rm}\fi
\def\tenpoint{\def\rm{\fam0\tenrm}
\textfont0=\tenrm \scriptfont0=\sevenrm \scriptscriptfont0=\fiverm
\textfont1=\teni  \scriptfont1=\seveni  \scriptscriptfont1=\fivei
\textfont2=\tensy \scriptfont2=\sevensy \scriptscriptfont2=\fivesy
\textfont\itfam=\tenit \def\it{\fam\itfam\tenit}\def\footnotefont{\ninepoint}%
\textfont\bffam=\tenbf \def\bf{\fam\bffam\tenbf}\def\sl{\fam\slfam\tensl}\rm}
\font\ninerm=cmr9 \font\sixrm=cmr6 \font\ninei=cmmi9 \font\sixi=cmmi6
\font\ninesy=cmsy9 \font\sixsy=cmsy6 \font\ninebf=cmbx9
\font\nineit=cmti9 \font\ninesl=cmsl9 \skewchar\ninei='177
\skewchar\sixi='177 \skewchar\ninesy='60 \skewchar\sixsy='60
\def\ninepoint{\def\rm{\fam0\ninerm}
\textfont0=\ninerm \scriptfont0=\sixrm \scriptscriptfont0=\fiverm
\textfont1=\ninei \scriptfont1=\sixi \scriptscriptfont1=\fivei
\textfont2=\ninesy \scriptfont2=\sixsy \scriptscriptfont2=\fivesy
\textfont\itfam=\ninei \def\it{\fam\itfam\nineit}\def\sl{\fam\slfam\ninesl}%
\textfont\bffam=\ninebf \def\bf{\fam\bffam\ninebf}\rm}
%
%

\hyphenation{anom-aly anom-alies coun-ter-term coun-ter-terms}
\def\inv{^{\raise.15ex\hbox{${\scriptscriptstyle -}$}\kern-.05em 1}}

\def\Dsl{\,\raise.15ex\hbox{/}\mkern-13.5mu D} 
\def\dsl{\raise.15ex\hbox{/}\kern-.57em\partial}

\def\lspace{\ifx\answ\bigans{}\else\qquad\fi}
\def\lbspace{\ifx\answ\bigans{}\else\hskip-.2in\fi} 
\def\boxeqn#1{\vcenter{\vbox{\hrule\hbox{\vrule\kern3pt\vbox{\kern3pt
	\hbox{${\displaystyle #1}$}\kern3pt}\kern3pt\vrule}\hrule}}}
\def\mbox#1#2{\vcenter{\hrule \hbox{\vrule height#2in
		\kern#1in \vrule} \hrule}}  
%

\def\darr#1{\raise1.5ex\hbox{$\leftrightarrow$}\mkern-16.5mu #1}

\def\roughly#1{\raise.3ex\hbox{$#1$\kern-.75em\lower1ex\hbox{$\sim$}}}

\def\smallfig#1#2#3{\DefWarn#1\xdef#1{fig.~\the\figno}
\writedef{#1\leftbracket fig.\noexpand~\the\figno}%
\figinsert\figin{\centerline{#3}}\medskip\centerline{\vbox{
\baselineskip12pt\advance\hsize by -1truein
\noindent\footnotefont{\bf Fig.~\the\figno:} #2}}
\endinsert\global\advance\figno by1}

\def\bb{
\font\tenmsb=msbm10
\font\sevenmsb=msbm7
\font\fivemsb=msbm5
\textfont1=\tenmsb
\scriptfont1=\sevenmsb
\scriptscriptfont1=\fivemsb
}

\input amssym

%
%
\ifx\pdfoutput\undefined
\input epsf
\def\fig#1{\epsfbox{#1.eps}}
\def\figscale#1#2{\epsfxsize=#2\epsfbox{#1.eps}}
%
%
\else
\def\fig#1{\pdfximage {#1.pdf}\pdfrefximage\pdflastximage}
\def\figscale#1#2{\pdfximage width#2 {#1.pdf}\pdfrefximage\pdflastximage}
\fi

\def\IZ{\relax\ifmmode\mathchoice
{\hbox{\cmss Z\kern-.4em Z}}{\hbox{\cmss Z\kern-.4em Z}} {\lower.9pt\hbox{\cmsss Z\kern-.4em Z}}
{\lower1.2pt\hbox{\cmsss Z\kern-.4em Z}}\else{\cmss Z\kern-.4em Z}\fi}

\newif\ifdraft\draftfalse
\newif\ifinter\interfalse
\ifdraft\draftmode\else\interfalse\fi
\def\journal#1&#2(#3){\unskip, \sl #1\ \bf #2 \rm(19#3) }
\def\andjournal#1&#2(#3){\sl #1~\bf #2 \rm (19#3) }

\def\frac#1#2{{#1\over#2}}

\def\inbar{\,\vrule height1.5ex width.4pt depth0pt}
\def\IC{\relax\hbox{$\inbar\kern-.3em{\rm C}$}}
\def\IR{\relax{\rm I\kern-.18em R}}
\def\IP{\relax{\rm I\kern-.18em P}}
\def\Z{{\bf Z}}

%
%


%
\catcode`\@=11
\def\slash#1{\mathord{\mathpalette\c@ncel{#1}}}
\overfullrule=0pt

\def\underrel#1\over#2{\mathrel{\mathop{\kern\z@#1}\limits_{#2}}}

\catcode`\@=12


%

\def\det{{\rm det}}

\def\det{{\rm det}}


\def\[{[}
\def\]{]}

\def\comment#1{ }

%
\def\draftnote#1{\ifdraft{\baselineskip2ex
                 \vbox{\kern1em\hrule\hbox{\vrule\kern1em\vbox{\kern1ex
                 \noindent \underbar{NOTE}: #1
             \vskip1ex}\kern1em\vrule}\hrule}}\fi}
\def\internote#1{\ifinter{\baselineskip2ex
                 \vbox{\kern1em\hrule\hbox{\vrule\kern1em\vbox{\kern1ex
                 \noindent \underbar{Internal Note}: #1
             \vskip1ex}\kern1em\vrule}\hrule}}\fi}

%
%



%
%
%
%

%

\def\inv{^{-1}}


\def\1{{\ds 1}}

\def\Z{\hbox{$\bb Z$}}

%
\def\draftnote#1{\ifdraft{\baselineskip2ex
                 \vbox{\kern1em\hrule\hbox{\vrule\kern1em\vbox{\kern1ex
                 \noindent \underbar{NOTE}: #1
             \vskip1ex}\kern1em\vrule}\hrule}}\fi}
\def\internote#1{\ifinter{\baselineskip2ex
                 \vbox{\kern1em\hrule\hbox{\vrule\kern1em\vbox{\kern1ex
                 \noindent \underbar{Internal Note}: #1
             \vskip1ex}\kern1em\vrule}\hrule}}\fi}

%
%



%
%
%
%

%

\def\inv{^{-1}}


\def\1{{\ds 1}}

\def\Z{\hbox{$\bb Z$}}

\def\S{\hbox{$\bb S$}}

\lref\NiarchosAH{
  V.~Niarchos,
  ``Seiberg dualities and the 3d/4d connection,''
JHEP {\bf 1207}, 075 (2012).
[arXiv:1205.2086 [hep-th]].
}

\lref\AharonyGP{
  O.~Aharony,
  ``IR duality in d = 3 N=2 supersymmetric USp(2N(c)) and U(N(c)) gauge theories,''
Phys.\ Lett.\ B {\bf 404}, 71 (1997).
[hep-th/9703215].
}

\lref\AffleckAS{
  I.~Affleck, J.~A.~Harvey and E.~Witten,
  ``Instantons and (Super)Symmetry Breaking in (2+1)-Dimensions,''
Nucl.\ Phys.\ B {\bf 206}, 413 (1982)..
}

\lref\IntriligatorID{
  K.~A.~Intriligator and N.~Seiberg,
  ``Duality, monopoles, dyons, confinement and oblique confinement in supersymmetric SO(N(c)) gauge theories,''
Nucl.\ Phys.\ B {\bf 444}, 125 (1995).
[hep-th/9503179].
}

\lref\PasquettiFJ{
  S.~Pasquetti,
  ``Factorisation of N = 2 Theories on the Squashed 3-Sphere,''
JHEP {\bf 1204}, 120 (2012).
[arXiv:1111.6905 [hep-th]].
}

\lref\BeemMB{
  C.~Beem, T.~Dimofte and S.~Pasquetti,
  ``Holomorphic Blocks in Three Dimensions,''
[arXiv:1211.1986 [hep-th]].
}

\lref\SeibergPQ{
  N.~Seiberg,
  ``Electric - magnetic duality in supersymmetric nonAbelian gauge theories,''
Nucl.\ Phys.\ B {\bf 435}, 129 (1995).
[hep-th/9411149].
}

\lref\AharonyBX{
  O.~Aharony, A.~Hanany, K.~A.~Intriligator, N.~Seiberg and M.~J.~Strassler,
  ``Aspects of N=2 supersymmetric gauge theories in three-dimensions,''
Nucl.\ Phys.\ B {\bf 499}, 67 (1997).
[hep-th/9703110].
}

\lref\IntriligatorNE{
  K.~A.~Intriligator and P.~Pouliot,
  ``Exact superpotentials, quantum vacua and duality in supersymmetric SP(N(c)) gauge theories,''
Phys.\ Lett.\ B {\bf 353}, 471 (1995).
[hep-th/9505006].
}

\lref\KarchUX{
  A.~Karch,
  ``Seiberg duality in three-dimensions,''
Phys.\ Lett.\ B {\bf 405}, 79 (1997).
[hep-th/9703172].
}

\lref\SafdiRE{
  B.~R.~Safdi, I.~R.~Klebanov and J.~Lee,
  ``A Crack in the Conformal Window,''
[arXiv:1212.4502 [hep-th]].
}

\lref\SchweigertTG{
  C.~Schweigert,
  ``On moduli spaces of flat connections with nonsimply connected structure group,''
Nucl.\ Phys.\ B {\bf 492}, 743 (1997).
[hep-th/9611092].
}

\lref\GiveonZN{
  A.~Giveon and D.~Kutasov,
  ``Seiberg Duality in Chern-Simons Theory,''
Nucl.\ Phys.\ B {\bf 812}, 1 (2009).
[arXiv:0808.0360 [hep-th]].
}

\lref\Spiridonov{
  Spiridonov, V.~P.,
  ``Aspects of elliptic hypergeometric functions,''
[arXiv:1307.2876 [math.CA]].
}

\lref\GaiottoBE{
  D.~Gaiotto, G.~W.~Moore and A.~Neitzke,
  ``Framed BPS States,''
[arXiv:1006.0146 [hep-th]].
}

\lref\AldayRS{
  L.~F.~Alday, M.~Bullimore and M.~Fluder,
  ``On S-duality of the Superconformal Index on Lens Spaces and 2d TQFT,''
JHEP {\bf 1305}, 122 (2013).
[arXiv:1301.7486 [hep-th]].
}

\lref\RazamatJXA{
  S.~S.~Razamat and M.~Yamazaki,
  ``S-duality and the N=2 Lens Space Index,''
[arXiv:1306.1543 [hep-th]].
}

\lref\NiarchosAH{
  V.~Niarchos,
  ``Seiberg dualities and the 3d/4d connection,''
JHEP {\bf 1207}, 075 (2012).
[arXiv:1205.2086 [hep-th]].
}

\lref\almost{
  A.~Borel, R.~Friedman, J.~W.~Morgan,
  ``Almost commuting elements in compact Lie groups,''
arXiv:math/9907007.
}

\lref\KapustinJM{
  A.~Kapustin and B.~Willett,
  ``Generalized Superconformal Index for Three Dimensional Field Theories,''
[arXiv:1106.2484 [hep-th]].
}

\lref\AharonyGP{
  O.~Aharony,
  ``IR duality in d = 3 N=2 supersymmetric USp(2N(c)) and U(N(c)) gauge theories,''
Phys.\ Lett.\ B {\bf 404}, 71 (1997).
[hep-th/9703215].
}

\lref\FestucciaWS{
  G.~Festuccia and N.~Seiberg,
  ``Rigid Supersymmetric Theories in Curved Superspace,''
JHEP {\bf 1106}, 114 (2011).
[arXiv:1105.0689 [hep-th]].
}

\lref\RomelsbergerEG{
  C.~Romelsberger,
  ``Counting chiral primaries in N = 1, d=4 superconformal field theories,''
Nucl.\ Phys.\ B {\bf 747}, 329 (2006).
[hep-th/0510060].
}

\lref\KapustinKZ{
  A.~Kapustin, B.~Willett and I.~Yaakov,
  ``Exact Results for Wilson Loops in Superconformal Chern-Simons Theories with Matter,''
JHEP {\bf 1003}, 089 (2010).
[arXiv:0909.4559 [hep-th]].
}

\lref\DolanQI{
  F.~A.~Dolan and H.~Osborn,
  ``Applications of the Superconformal Index for Protected Operators and q-Hypergeometric Identities to N=1 Dual Theories,''
Nucl.\ Phys.\ B {\bf 818}, 137 (2009).
[arXiv:0801.4947 [hep-th]].
}

\lref\GaddeIA{
  A.~Gadde and W.~Yan,
  ``Reducing the 4d Index to the $S^3$ Partition Function,''
JHEP {\bf 1212}, 003 (2012).
[arXiv:1104.2592 [hep-th]].
}

\lref\DolanRP{
  F.~A.~H.~Dolan, V.~P.~Spiridonov and G.~S.~Vartanov,
  ``From 4d superconformal indices to 3d partition functions,''
Phys.\ Lett.\ B {\bf 704}, 234 (2011).
[arXiv:1104.1787 [hep-th]].
}

\lref\ImamuraUW{
  Y.~Imamura,
 ``Relation between the 4d superconformal index and the $S^3$ partition function,''
JHEP {\bf 1109}, 133 (2011).
[arXiv:1104.4482 [hep-th]].
}

\lref\BeemYN{
  C.~Beem and A.~Gadde,
  ``The $N=1$ superconformal index for class $S$ fixed points,''
JHEP {\bf 1404}, 036 (2014).
[arXiv:1212.1467 [hep-th]].
}

\lref\HamaEA{
  N.~Hama, K.~Hosomichi and S.~Lee,
  ``SUSY Gauge Theories on Squashed Three-Spheres,''
JHEP {\bf 1105}, 014 (2011).
[arXiv:1102.4716 [hep-th]].
}

\lref\GaddeEN{
  A.~Gadde, L.~Rastelli, S.~S.~Razamat and W.~Yan,
  ``On the Superconformal Index of N=1 IR Fixed Points: A Holographic Check,''
JHEP {\bf 1103}, 041 (2011).
[arXiv:1011.5278 [hep-th]].
}

\lref\EagerHX{
  R.~Eager, J.~Schmude and Y.~Tachikawa,
  ``Superconformal Indices, Sasaki-Einstein Manifolds, and Cyclic Homologies,''
[arXiv:1207.0573 [hep-th]].
}

\lref\AffleckAS{
  I.~Affleck, J.~A.~Harvey and E.~Witten,
  ``Instantons and (Super)Symmetry Breaking in (2+1)-Dimensions,''
Nucl.\ Phys.\ B {\bf 206}, 413 (1982)..
}

\lref\SeibergPQ{
  N.~Seiberg,
  ``Electric - magnetic duality in supersymmetric nonAbelian gauge theories,''
Nucl.\ Phys.\ B {\bf 435}, 129 (1995).
[hep-th/9411149].
}

\lref\BahDG{
  I.~Bah, C.~Beem, N.~Bobev and B.~Wecht,
  ``Four-Dimensional SCFTs from M5-Branes,''
JHEP {\bf 1206}, 005 (2012).
[arXiv:1203.0303 [hep-th]].
}

\lref\debult{
  F.~van~de~Bult,
  ``Hyperbolic Hypergeometric Functions,''
University of Amsterdam Ph.D. thesis
}

\lref\Shamirthesis{
  I.~Shamir,
  ``Aspects of three dimensional Seiberg duality,''
  M. Sc. thesis submitted to the Weizmann Institute of Science, April 2010.
  }

\lref\slthreeZ{
  J.~Felder, A.~Varchenko,
  ``The elliptic gamma function and $SL(3,Z) \times Z^3$,'' $\;\;$
[arXiv:math/0001184].
}

\lref\BeniniNC{
  F.~Benini, T.~Nishioka and M.~Yamazaki,
  ``4d Index to 3d Index and 2d TQFT,''
Phys.\ Rev.\ D {\bf 86}, 065015 (2012).
[arXiv:1109.0283 [hep-th]].
}

\lref\GaiottoWE{
  D.~Gaiotto,
  ``N=2 dualities,''
  JHEP {\bf 1208}, 034 (2012).
  [arXiv:0904.2715 [hep-th]].
}

\lref\BeniniNDA{
  F.~Benini, R.~Eager, K.~Hori and Y.~Tachikawa,
  ``Elliptic genera of two-dimensional N=2 gauge theories with rank-one gauge groups,''
Lett.\ Math.\ Phys.\  {\bf 104}, 465 (2014).
[arXiv:1305.0533 [hep-th]].
}

\lref\SpiridonovZA{
  V.~P.~Spiridonov and G.~S.~Vartanov,
  ``Elliptic Hypergeometry of Supersymmetric Dualities,''
Commun.\ Math.\ Phys.\  {\bf 304}, 797 (2011).
[arXiv:0910.5944 [hep-th]].
}

\lref\BeniniMF{
  F.~Benini, C.~Closset and S.~Cremonesi,
  ``Comments on 3d Seiberg-like dualities,''
JHEP {\bf 1110}, 075 (2011).
[arXiv:1108.5373 [hep-th]].
}

\lref\ClossetVP{
  C.~Closset, T.~T.~Dumitrescu, G.~Festuccia, Z.~Komargodski and N.~Seiberg,
  ``Comments on Chern-Simons Contact Terms in Three Dimensions,''
JHEP {\bf 1209}, 091 (2012).
[arXiv:1206.5218 [hep-th]].
}

\lref\SpiridonovHF{
  V.~P.~Spiridonov and G.~S.~Vartanov,
  ``Elliptic hypergeometry of supersymmetric dualities II. Orthogonal groups, knots, and vortices,''
[arXiv:1107.5788 [hep-th]].
}

\lref\SpiridonovWW{
  V.~P.~Spiridonov and G.~S.~Vartanov,
  ``Elliptic hypergeometric integrals and 't Hooft anomaly matching conditions,''
JHEP {\bf 1206}, 016 (2012).
[arXiv:1203.5677 [hep-th]].
}

\lref\DimoftePY{
  T.~Dimofte, D.~Gaiotto and S.~Gukov,
  ``3-Manifolds and 3d Indices,''
[arXiv:1112.5179 [hep-th]].
}

\lref\KimWB{
  S.~Kim,
  ``The Complete superconformal index for N=6 Chern-Simons theory,''
Nucl.\ Phys.\ B {\bf 821}, 241 (2009), [Erratum-ibid.\ B {\bf 864}, 884 (2012)].
[arXiv:0903.4172 [hep-th]].
}

\lref\WillettGP{
  B.~Willett and I.~Yaakov,
  ``N=2 Dualities and Z Extremization in Three Dimensions,''
[arXiv:1104.0487 [hep-th]].
}

\lref\HeckmanBFA{
  J.~J.~Heckman, D.~R.~Morrison, T.~Rudelius and C.~Vafa,
  ``Atomic Classification of 6D SCFTs,''
Fortsch.\ Phys.\  {\bf 63}, 468 (2015).
[arXiv:1502.05405 [hep-th]].
}

\lref\ImamuraSU{
  Y.~Imamura and S.~Yokoyama,
  ``Index for three dimensional superconformal field theories with general R-charge assignments,''
JHEP {\bf 1104}, 007 (2011).
[arXiv:1101.0557 [hep-th]].
}

\lref\FreedYA{
  D.~S.~Freed, G.~W.~Moore and G.~Segal,
  ``The Uncertainty of Fluxes,''
Commun.\ Math.\ Phys.\  {\bf 271}, 247 (2007).
[hep-th/0605198].
}

\lref\HwangQT{
  C.~Hwang, H.~Kim, K.~-J.~Park and J.~Park,
  ``Index computation for 3d Chern-Simons matter theory: test of Seiberg-like duality,''
JHEP {\bf 1109}, 037 (2011).
[arXiv:1107.4942 [hep-th]].
}

\lref\GreenDA{
  D.~Green, Z.~Komargodski, N.~Seiberg, Y.~Tachikawa and B.~Wecht,
  ``Exactly Marginal Deformations and Global Symmetries,''
JHEP {\bf 1006}, 106 (2010).
[arXiv:1005.3546 [hep-th]].
}

\lref\GaiottoXA{
  D.~Gaiotto, L.~Rastelli and S.~S.~Razamat,
  ``Bootstrapping the superconformal index with surface defects,''
[arXiv:1207.3577 [hep-th]].
}

\lref\IntriligatorID{
  K.~A.~Intriligator and N.~Seiberg,
  ``Duality, monopoles, dyons, confinement and oblique confinement in supersymmetric SO(N(c)) gauge theories,''
Nucl.\ Phys.\ B {\bf 444}, 125 (1995).
[hep-th/9503179].
}

\lref\SeibergNZ{
  N.~Seiberg and E.~Witten,
  ``Gauge dynamics and compactification to three-dimensions,''
In *Saclay 1996, The mathematical beauty of physics* 333-366.
[hep-th/9607163].
}

\lref\KinneyEJ{
  J.~Kinney, J.~M.~Maldacena, S.~Minwalla and S.~Raju,
  ``An Index for 4 dimensional super conformal theories,''
  Commun.\ Math.\ Phys.\  {\bf 275}, 209 (2007).
  [hep-th/0510251].
}

\lref\NakayamaUR{
  Y.~Nakayama,
  ``Index for supergravity on AdS(5) x T**1,1 and conifold gauge theory,''
Nucl.\ Phys.\ B {\bf 755}, 295 (2006).
[hep-th/0602284].
}

\lref\GaddeKB{
  A.~Gadde, E.~Pomoni, L.~Rastelli and S.~S.~Razamat,
  ``S-duality and 2d Topological QFT,''
JHEP {\bf 1003}, 032 (2010).
[arXiv:0910.2225 [hep-th]].
}

\lref\GaddeTE{
  A.~Gadde, L.~Rastelli, S.~S.~Razamat and W.~Yan,
  ``The Superconformal Index of the $E_6$ SCFT,''
JHEP {\bf 1008}, 107 (2010).
[arXiv:1003.4244 [hep-th]].
}

\lref\AharonyCI{
  O.~Aharony and I.~Shamir,
  ``On $O(N_c)$ d=3 N=2 supersymmetric QCD Theories,''
JHEP {\bf 1112}, 043 (2011).
[arXiv:1109.5081 [hep-th]].
}

\lref\GiveonSR{
  A.~Giveon and D.~Kutasov,
  ``Brane dynamics and gauge theory,''
Rev.\ Mod.\ Phys.\  {\bf 71}, 983 (1999).
[hep-th/9802067].
}

\lref\SpiridonovQV{
  V.~P.~Spiridonov and G.~S.~Vartanov,
  ``Superconformal indices of ${\cal N}=4$ SYM field theories,''
Lett.\ Math.\ Phys.\  {\bf 100}, 97 (2012).
[arXiv:1005.4196 [hep-th]].
}
\lref\GaddeUV{
  A.~Gadde, L.~Rastelli, S.~S.~Razamat and W.~Yan,
  ``Gauge Theories and Macdonald Polynomials,''
Commun.\ Math.\ Phys.\  {\bf 319}, 147 (2013).
[arXiv:1110.3740 [hep-th]].
}
\lref\KapustinGH{
  A.~Kapustin,
  ``Seiberg-like duality in three dimensions for orthogonal gauge groups,''
[arXiv:1104.0466 [hep-th]].
}

\lref\orthogpaper{O. Aharony, S. S. Razamat, N.~Seiberg and B.~Willett, 
``3d dualities from 4d dualities for orthogonal groups,''
[arXiv:1307.0511 [hep-th]].
}

\lref\readinglines{
  O.~Aharony, N.~Seiberg and Y.~Tachikawa,
  ``Reading between the lines of four-dimensional gauge theories,''
[arXiv:1305.0318 [hep-th]].
}

\lref\WittenNV{
  E.~Witten,
  ``Supersymmetric index in four-dimensional gauge theories,''
Adv.\ Theor.\ Math.\ Phys.\  {\bf 5}, 841 (2002).
[hep-th/0006010].
}

\lref\GaddeUV{
  A.~Gadde, L.~Rastelli, S.~S.~Razamat and W.~Yan,
  ``Gauge Theories and Macdonald Polynomials,''
Commun.\ Math.\ Phys.\  {\bf 319}, 147 (2013).
[arXiv:1110.3740 [hep-th]].
}

\lref\GaddeIK{
  A.~Gadde, L.~Rastelli, S.~S.~Razamat and W.~Yan,
  ``The 4d Superconformal Index from q-deformed 2d Yang-Mills,''
Phys.\ Rev.\ Lett.\  {\bf 106}, 241602 (2011).
[arXiv:1104.3850 [hep-th]].
}

\lref\GaiottoXA{
  D.~Gaiotto, L.~Rastelli and S.~S.~Razamat,
  ``Bootstrapping the superconformal index with surface defects,''
JHEP {\bf 1301}, 022 (2013).
[arXiv:1207.3577 [hep-th]].
}

\lref\GaiottoUQ{
  D.~Gaiotto and S.~S.~Razamat,
  ``Exceptional Indices,''
JHEP {\bf 1205}, 145 (2012).
[arXiv:1203.5517 [hep-th]].
}

\lref\RazamatUV{
  S.~S.~Razamat,
  ``On a modular property of N=2 superconformal theories in four dimensions,''
JHEP {\bf 1210}, 191 (2012).
[arXiv:1208.5056 [hep-th]].
}

\lref\noumi{
  Y.~Komori, M.~Noumi, J.~Shiraishi,
  ``Kernel Functions for Difference Operators of Ruijsenaars Type and Their Applications,''
SIGMA 5 (2009), 054.
[arXiv:0812.0279 [math.QA]].
}

\lref\SpirWarnaar{
  V.~P.~Spiridonov and S.~O.~Warnaar,
  ``Inversions of integral operators and elliptic beta integrals on root systems,''
Adv. Math. 207 (2006), 91-132
[arXiv:math/0411044].
}

\lref\RazamatJXA{
  S.~S.~Razamat and M.~Yamazaki,
  ``S-duality and the N=2 Lens Space Index,''
[arXiv:1306.1543 [hep-th]].
}

\lref\RazamatOPA{
  S.~S.~Razamat and B.~Willett,
  ``Global Properties of Supersymmetric Theories and the Lens Space,''
[arXiv:1307.4381 [hep-th]].
}

\lref\GaddeTE{
  A.~Gadde, L.~Rastelli, S.~S.~Razamat and W.~Yan,
  ``The Superconformal Index of the $E_6$ SCFT,''
JHEP {\bf 1008}, 107 (2010).
[arXiv:1003.4244 [hep-th]].
}

\lref\deBult{
  F.~J.~van~de~Bult,
  ``An elliptic hypergeometric integral with $W(F_4)$ symmetry,''
The Ramanujan Journal, Volume 25, Issue 1 (2011)
[arXiv:0909.4793[math.CA]].
}

\lref\GaddeKB{
  A.~Gadde, E.~Pomoni, L.~Rastelli and S.~S.~Razamat,
  ``S-duality and 2d Topological QFT,''
JHEP {\bf 1003}, 032 (2010).
[arXiv:0910.2225 [hep-th]].
}

\lref\ArgyresCN{
  P.~C.~Argyres and N.~Seiberg,
  ``S-duality in N=2 supersymmetric gauge theories,''
JHEP {\bf 0712}, 088 (2007).
[arXiv:0711.0054 [hep-th]].
}

\lref\RastelliTBZ{
  L.~Rastelli and S.~S.~Razamat,
  ``The supersymmetric index in four dimensions,''
J.\ Phys.\ A {\bf 50}, no. 44, 443013 (2017).
[arXiv:1608.02965 [hep-th]].
}

\lref\SpirWarnaar{
  V.~P.~Spiridonov and S.~O.~Warnaar,
  ``Inversions of integral operators and elliptic beta integrals on root systems,''
Adv. Math. 207 (2006), 91-132
[arXiv:math/0411044].
}

\lref\GaiottoHG{
  D.~Gaiotto, G.~W.~Moore and A.~Neitzke,
  ``Wall-crossing, Hitchin Systems, and the WKB Approximation,''
[arXiv:0907.3987 [hep-th]].
}

\lref\DrukkerDGN{
  N.~Drukker, I.~Shamir and C.~Vergu,
  ``Defect multiplets of $ {\cal N}=1 $ supersymmetry in 4d,''
JHEP {\bf 1801}, 034 (2018).
[arXiv:1711.03455 [hep-th]].
}

\lref\RuijsenaarsVQ{
  S.~N.~M.~Ruijsenaars and H.~Schneider,
  ``A New Class Of Integrable Systems And Its Relation To Solitons,''
Annals Phys.\  {\bf 170}, 370 (1986).
}

\lref\RuijsenaarsPP{
  S.~N.~M.~Ruijsenaars,
  ``Complete Integrability Of Relativistic Calogero-moser Systems And Elliptic Function Identities,''
Commun.\ Math.\ Phys.\  {\bf 110}, 191 (1987).
}

\lref\HallnasNB{
  M.~Hallnas and S.~Ruijsenaars,
  ``Kernel functions and Baecklund transformations for relativistic Calogero-Moser and Toda systems,''
J.\ Math.\ Phys.\  {\bf 53}, 123512 (2012).
}

\lref\kernelA{
S.~Ruijsenaars,
  ``Elliptic integrable systems of Calogero-Moser type: Some new results on joint eigenfunctions'', in Proceedings of the 2004 Kyoto Workshop on "Elliptic integrable systems", (M. Noumi, K. Takasaki, Eds.), Rokko Lectures in Math., no. 18, Dept. of Math., Kobe Univ.
}

\lref\ellRSreview{
Y.~Komori and S.~Ruijsenaars,
  ``Elliptic integrable systems of Calogero-Moser type: A survey'', in Proceedings of the 2004 Kyoto Workshop on "Elliptic integrable systems", (M. Noumi, K. Takasaki, Eds.), Rokko Lectures in Math., no. 18, Dept. of Math., Kobe Univ.
}

\lref\langmann{
E.~Langmann,
  ``An explicit solution of the (quantum) elliptic Calogero-Sutherland model'', [arXiv:math-ph/0407050].
}

\lref\TachikawaWI{
  Y.~Tachikawa,
  ``4d partition function on $S^1 \times S^3$ and 2d Yang-Mills with nonzero area,''
PTEP {\bf 2013}, 013B01 (2013).
[arXiv:1207.3497 [hep-th]].
}

\lref\ItoFPL{
  Y.~Ito and Y.~Yoshida,
  ``Superconformal index with surface defects for class ${\cal S}_k$,''
[arXiv:1606.01653 [hep-th]].
}

\lref\MinahanFG{
  J.~A.~Minahan and D.~Nemeschansky,
  ``An N=2 superconformal fixed point with E(6) global symmetry,''
Nucl.\ Phys.\ B {\bf 482}, 142 (1996).
[hep-th/9608047].
}

\lref\AldayKDA{
  L.~F.~Alday, M.~Bullimore, M.~Fluder and L.~Hollands,
  ``Surface defects, the superconformal index and q-deformed Yang-Mills,''
[arXiv:1303.4460 [hep-th]].
}

\lref\NekrasovCZ{
  N.~Nekrasov,
  ``Five dimensional gauge theories and relativistic integrable systems,''
Nucl.\ Phys.\ B {\bf 531}, 323 (1998).
[hep-th/9609219].
}

\lref\FukudaJR{
  Y.~Fukuda, T.~Kawano and N.~Matsumiya,
  ``5D SYM and 2D q-Deformed YM,''
Nucl.\ Phys.\ B {\bf 869}, 493 (2013).
[arXiv:1210.2855 [hep-th]].
}

\lref\XieHS{
  D.~Xie,
  ``General Argyres-Douglas Theory,''
JHEP {\bf 1301}, 100 (2013).
[arXiv:1204.2270 [hep-th]].
}

\lref\DrukkerSR{
  N.~Drukker, T.~Okuda and F.~Passerini,
  ``Exact results for vortex loop operators in 3d supersymmetric theories,''
[arXiv:1211.3409 [hep-th]].
}

\lref\qinteg{
  M.~Rahman, A.~Verma,
  ``A q-integral representation of Rogers' q-ultraspherical polynomials and some applications,''
Constructive Approximation
1986, Volume 2, Issue 1.
}

\lref\qintegOK{
  A.~Okounkov,
  ``(Shifted) Macdonald Polynomials: q-Integral Representation and Combinatorial Formula,''
Compositio Mathematica
June 1998, Volume 112, Issue 2. 
[arXiv:q-alg/9605013].
}

\lref\JeffersonAHM{
  P.~Jefferson, H.~C.~Kim, C.~Vafa and G.~Zafrir,
  ``Towards Classification of 5d SCFTs: Single Gauge Node,''
[arXiv:1705.05836 [hep-th]].
}

\lref\MaruyoshiCAF{
  K.~Maruyoshi and J.~Yagi,
  ``Surface defects as transfer matrices,''
PTEP {\bf 2016}, no. 11, 113B01 (2016).
[arXiv:1606.01041 [hep-th]].
}

\lref\macNest{
 H.~Awata, S.~Odake, J.~Shiraishi,
  ``Integral Representations of the Macdonald Symmetric Functions,''
Commun. Math. Phys. 179 (1996) 647.
[arXiv:q-alg/9506006].
}

\lref\AldayKDA{
  L.~F.~Alday, M.~Bullimore, M.~Fluder and L.~Hollands,
  ``Surface defects, the superconformal index and q-deformed Yang-Mills,''
JHEP {\bf 1310}, 018 (2013).
[arXiv:1303.4460 [hep-th]].
}

\lref\BeemYN{
  C.~Beem and A.~Gadde,
  ``The superconformal index of N=1 class S fixed points,''
[arXiv:1212.1467 [hep-th]].
}

\lref\GaddeFMA{
  A.~Gadde, K.~Maruyoshi, Y.~Tachikawa and W.~Yan,
  ``New N=1 Dualities,''
JHEP {\bf 1306}, 056 (2013).
[arXiv:1303.0836 [hep-th]].
}

 \lref\RazamatGRO{
  S.~S.~Razamat and G.~Zafrir,
  ``Compactification of 6d minimal SCFTs on Riemann surfaces,''
[arXiv:1806.09196 [hep-th]].
}

\lref\GorskyTN{
  A.~Gorsky,
  ``Dualities in integrable systems and N=2 SUSY theories,''
J.\ Phys.\ A {\bf 34}, 2389 (2001).
[hep-th/9911037].
}

\lref\vandijm{
J.~F.~ van Diejen, ``Integrability of difference Calogero --  
Moser systems,'' J. Math. Phys. 35 (1994), 2983 -- 3004 
}

\lref\FockAE{
  V.~Fock, A.~Gorsky, N.~	 and V.~Rubtsov,
  ``Duality in integrable systems and gauge theories,''
JHEP {\bf 0007}, 028 (2000).
[hep-th/9906235].
}

\lref\NazzalBRC{
  B.~Nazzal and S.~S.~Razamat,
  ``Surface defects in E-string compactifications and the van Diejen model,''
SIGMA {\bf 14}, 036 (2018).
[arXiv:1801.00960 [hep-th]].
}

\lref\CsakiCU{
  C.~Csaki, M.~Schmaltz, W.~Skiba and J.~Terning,
  ``Selfdual N=1 SUSY gauge theories,''
Phys.\ Rev.\ D {\bf 56}, 1228 (1997).
[hep-th/9701191].
}

\lref\GaiottoUSA{
  D.~Gaiotto and S.~S.~Razamat,
  ``$ {\cal N}=1$ theories of class $ {\cal{S}}_k $,''
JHEP {\bf 1507}, 073 (2015).
[arXiv:1503.05159 [hep-th]].
}

\lref\GorskyPE{
  A.~Gorsky and N.~Nekrasov,
  ``Hamiltonian systems of Calogero type and two-dimensional Yang-Mills theory,''
Nucl.\ Phys.\ B {\bf 414}, 213 (1994).
[hep-th/9304047].
}

\lref\SeibergQX{
  N.~Seiberg,
  ``Nontrivial fixed points of the renormalization group in six-dimensions,''
Phys.\ Lett.\ B {\bf 390}, 169 (1997).
[hep-th/9609161].
}

\lref\BershadskySB{
  M.~Bershadsky and C.~Vafa,
  ``Global anomalies and geometric engineering of critical theories in six-dimensions,''
[hep-th/9703167].
}

\lref\NekrasovRC{
  N.~A.~Nekrasov and S.~L.~Shatashvili,
  ``Quantization of Integrable Systems and Four Dimensional Gauge Theories,''
[arXiv:0908.4052 [hep-th]].
}

\lref\DonagiCF{
  R.~Donagi and E.~Witten,
  ``Supersymmetric Yang-Mills theory and integrable systems,''
Nucl.\ Phys.\ B {\bf 460}, 299 (1996).
[hep-th/9510101].
}

\lref\GaddeFTV{
  A.~Gadde and S.~Gukov,
  ``2d Index and Surface operators,''
JHEP {\bf 1403}, 080 (2014).
[arXiv:1305.0266 [hep-th]].
}

\lref\GorskyZQ{
  A.~Gorsky, I.~Krichever, A.~Marshakov, A.~Mironov and A.~Morozov,
  ``Integrability and Seiberg-Witten exact solution,''
Phys.\ Lett.\ B {\bf 355}, 466 (1995).
[hep-th/9505035].
}

\lref\debult{
F.~van de Bult, F.~J.,
``An elliptic hypergeometric integral with $W(F_4)$ symmetry'',
The Ramanujan Journal, V. 25, Issue 1 (2011).
[arXiv:0909.4793].
}

\lref\Rains{
E.~Rains,
``Transformations of elliptic hypergometric integrals'',
[math/0309252].
}

\Title{\vbox{\baselineskip12pt
}}
{\vbox{
\centerline{Flavored surface defects in $4d$ ${\cal N}=1$ SCFTs }
\vskip7pt 
\centerline{}
}
}

\centerline{Shlomo S. Razamat}
\bigskip
\centerline{{\it Physics Department, Technion, Haifa, Israel 32000}}
\vskip.2in \vskip.3in \centerline{\bf Abstract}

We discuss supersymmetric surface defects in compactifications of six dimensional minimal conformal matter of type $SU(3)$ and $SO(8)$ to four dimensions. 
The relevant field theories in four dimensions are ${\cal N}=1$ quiver gauge theories with $SU(3)$ and $SU(4)$ gauge groups respectively. 
The defects are engineered by giving space-time dependent vacuum expectation values to baryonic operators.
We find evidence that in the  case of $SU(3)$ minimal conformal matter the defects carry  $SU(2)$ flavor symmetry which is {\it not a symmetry} of the four dimensional model. 
The simplest case of a model in this class is $SU(3)$ SQCD with nine flavors and thus the results suggest that this admits natural  surface defects with $SU(2)$ flavor symmetry.
We analyze the defects using the superconformal index and derive analytic difference operators introducing the defects into the index computation. The duality properties of the four dimensional theories imply that the index of the models is a kernel function for such difference operators. In turn, checking the kernel property constitutes an independent check of the dualities and the dictionary between six dimensional compactifications and four dimensional models.

\vskip.2in

\noindent

\vfill

\Date{}

\newsec{Introduction}

Surface defects are interesting non-local observables in quantum field theories and they  have received some attention in recent years. There are various ways to introduce such defects into a $d$ dimensional model. For example, one can try and couple the degrees of freedom of a $d$ dimensional CFT to a two dimensional CFT, or define the defect by specifying  boundary conditions  of the $d$ dimensional theory supported on a two dimensional surface. Yet another way to introduce surface defects is by studying flows triggered by vacuum expectation values with a non trivial space-time profile. Such flow can lead to an IR CFT with some of the degrees of freedom localized to submanifolds of the $d$ dimensional spacetime where the vacuum expectation value has special properties.

Combining the latter approach with supersymmetry can lead to quantitative tools to study the defects. In  this brief note we will study defects in a simple class of ${\cal N}=1$ supersymmetric field theories in four dimensions. This class of theories can be engineered as compactification of  six dimensional minimal conformal matter of type $SU(3)$ and $SO(8)$ \SeibergQX\BershadskySB\foot{The minimal conformal matter is described on the tensor branch as YM with single  simple gauge group  factor and single tensor multiplet with no matter. This set   of models is a subset of the so called non-higgsable cluster theories \HeckmanBFA\ in six dimensions.}  on a Riemann surface with punctures. The  theories in four dimensions obtained in such compactifications were identified in \RazamatGRO\ as certain gauge theories with $SU(3)$ and $SU(4)$ gauge groups respectively.
 We will observe a simple manifestation of an interesting phenomenon. In the class of theories obtained from minimal conformal matter of type  $SU(3)$  there are supersymmetric surface defects  that  carry degrees of freedom charged under  symmetry which is not a symmetry of the theory in the bulk.  We will argue for this by engineering the defects with the RG flow construction starting from a theory which has extra symmetry. Turning on constant vacuum expectation values some of the extra symmetry is explicitly broken and the rest does not appear in the IR fixed point as the fields charged under it become massive. However, when we will turn on space-time dependent vacuum expectation values, some remnants of the massive fields will survive on two dimensional subspace where the vacuum expectation value vanishes. This will thus produce interesting defects with additional symmetry which is naively surprizing  from the point of view of the bulk model. An interesting question, which we leave for future research, is to understand whether our defects can be engineered in other manner, say by coupling the four dimensional models to two dimensional theories as in \GaddeFTV\DrukkerDGN.

The fact that the theories we will consider have a geometric interpretation will have a mathematical implication. An interesting case here is when the six dimensional model we start with has an effective description as a five dimensional gauge theory when compactified on a circle. The  theories in four dimensions, in addition to the symmetry of the six dimensional model preserved in the compactification, also have factors of global symmetry associated to the punctures and being a subgroup of the five dimensional gauge symmetry. The theories that we will study are of this type. In particular they have a description in five dimensions as $SU(3)$ Chern-Simons model for the (twisted) compactification of $SU(3)$ minimal conformal matter, and $SU(4)$ Chern-Simons model for the (twisted) compactification of the $SO(8)$ minimal conformal matter \JeffersonAHM\ (see  discussion in \RazamatGRO).  We will compute the supersymmetric index of the theories in presence of surface defects. It is  given by certain analytic difference operators acting on the index of the theory without the defect. The difference operator one obtains is a Hamiltonian of a relativistic quantum mechanical model which might be associated to the five dimensional gauge theory \NekrasovCZ\NekrasovRC. In the well studied case of class ${\cal S}$ \GaiottoWE\GaiottoHG\ the relevant model is Ruijsenaars-Schneider integrable system \RuijsenaarsVQ\ as observed by Nekrasov in \NekrasovCZ\ long time ago and obtained in the context of index computations in \GaiottoXA. Other examples are the van Diejen model \vandijm\ for the E-string \NazzalBRC\ and some more intricate systems for class ${\cal S}_k$ models \GaiottoUSA\MaruyoshiCAF\ItoFPL. We will identify the relevant quantum mechanical systems in the two  classes of theories we study. We verify various properties such Hamiltonians have to satisfy following the conjectured map between compactifications and four dimensional theories of \RazamatGRO. This constitutes additional check of the
 conjectures.

We organize the paper as follows. In section two we will consider defects in theories obtained by compactification of the minimal conformal matter of type $SU(3)$. We will first discuss the basic physical considerations related to the defects and review the essentials of the field theories in four dimensions obtained in this compactification. We will observe that the defect might carry some symmetry. Next we will discuss the index in presence of the defects, and observe that indeed the defects carry  an $SU(2)$ symmetry, and will derive the quantum mechanical model associated to this construction. In section three we will consider defects in theories obtained by compactifications of minimal conformal matter of type $SO(8)$. In Appendix we summarize some technicalities.

\

\newsec{Case of the $SU(3)$ minimal conformal matter}

Let us first discuss the construction of the defect in the models we are going to consider from a simple perspective. The theories we consider are constructed by combining two ${\cal N}=1$ superconformal CFTs, $T_1$ and $T_2$, which have a factor of $SU(3)$ flavor symmetry each. We combine the models by coupling them to a triplet of bifundamental fields $Q_i$, a trifundamental, by gauging the two $SU(3)$ symmetries. In our constructions the only non anomalous flavor  symmetry the $Q_i$ are charged under is an $SU(3)_f$ rotating the three fields, and the superconformal R symmetry is $2/3$.  We then consider giving a vacuum expectation value to the baryon,

\eqn\lejrhgrkelw{
\det \,Q_1= \epsilon_{lmk}\epsilon^{ijn} (Q_1)_j^l(Q_1)_i^k (Q_1)_n^m\,.
} We can choose baryons built from $Q_2$ and $Q_3$ with equivalent results.
 This vacuum expectation value breaks explicitly the $SU(3)_f$ symmetry to $SU(2)_f$. The two gauge symmetries are Higgsed down to a diagonal combination with the $Q_1$ and $Q_2$ fields transforming in the adjoint (plus singlet) of the diagonal $SU(3)$ gauge symmetry and acquiring a mass term $Q_1Q_2$. Note that if there would have been a $U(1)$ symmetry under which $Q_i$ are charged a mass term would not be generated and the theory would have the $SU(2)_f$ symmetry in the IR. In the IR the theory one obtains is just the two models $T_1$ and $T_2$ combined by gauging a diagonal combination of the two $SU(3)$ symmetries, for illustration see Figure one. Note that although we did not break the $SU(2)_f$ symmetry, nothing in the IR is charged under it. 

\

 \

\centerline{\figscale{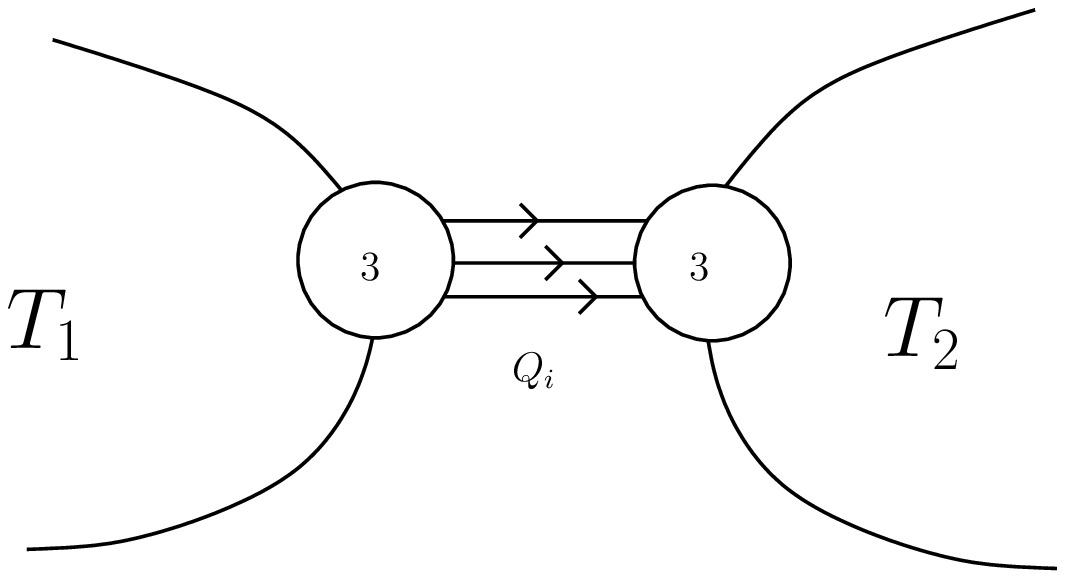}{2.4in} $\;\;\;\;\;\qquad \qquad $  \figscale{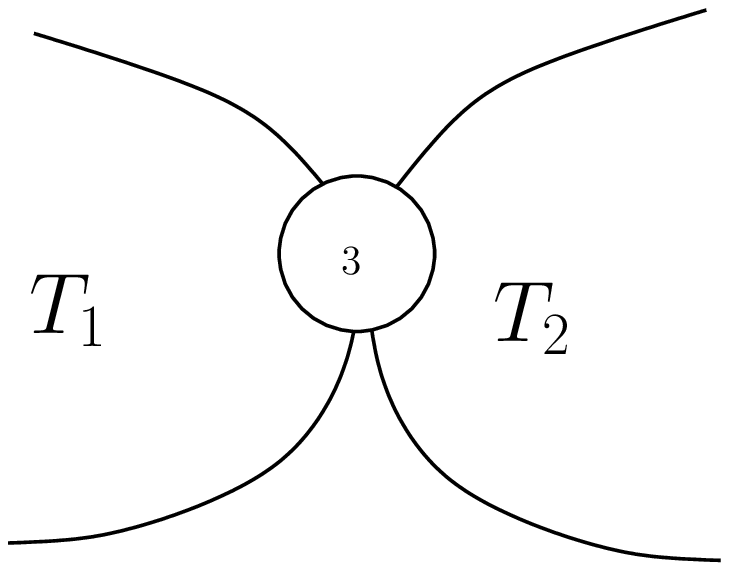}{1.6in}}
\medskip\centerline{\vbox{
\baselineskip12pt\advance\hsize by -1truein
\noindent\footnotefont{\bf Fig.~1:} Two theories $T_1$ and $T_2$ combined by gauging two $SU(3)$ symmetries connected by three bifundamental chiral fields $Q_i$.  We assume that there is no $U(1)$ symmetry under which $Q_i$ are charged.
We give vacuum expectation value to a baryon $ \det  Q_1$. The theory in the IR is the two theories $T_2$ and $T_1$  glued by gauging diagonal combination of two $SU(3)$ symmetries. If we give a vacuum expectation value to derivative of $\det Q_1$ we obtain surface defect in the theory on the right with some of the states charged under $SU(2)_f$ localized to the defect.}}

\

Next we consider the same construction but we give a space time dependent vacuum expectation value.  The logic is detailed in \GaiottoXA. 
We turn on a constant expectation value for $\partial \,\,\det \; Q_1$  where $\partial $ is a  derivative  in some plane in four  dimensions. For example, the plane is parametrized by complex coordinate $z$ and we take a holomorphic derivative.  Away from the locus $z=0$ we have a non vanishing vacuum expectation value and flow to the same theory as above. However at $z=0$ the vacuum expectation value vanishes and we might have some additional degrees of freedom localize on the two dimensional surface orthogonal to the complex plane parametrized by $z$. In particular there is no reason to expect that there is no remnant of the fields charged under the $SU(2)_f$ symmetry not broken by the vacuum expectation value. We will indeed see explicitly in the index computation below that this is the case and the defects carry degrees of freedom charged under the symmetry $SU(2)_f$ in the IR.

Before turning  to the index computation let us briefly review the construction of the theories $T_1$ and $T_2$ \RazamatGRO. We refer the reader for details to this reference. The claim is that the theories in four dimensions corresponding to (twisted) compactifications of the minimal conformal matter of type $SU(3)$ on a general Riemann surface are constructed from the  two simple blocks depicted in Figure two.
 
 \
 
 \centerline{\figscale{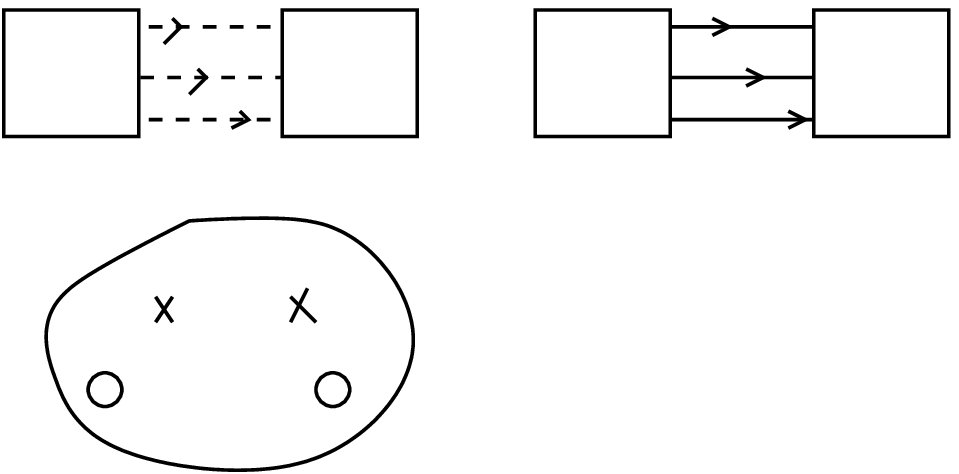}{2.6in}}
\medskip\centerline{\vbox{
\baselineskip12pt\advance\hsize by -1truein
\noindent\footnotefont{\bf Fig.~2:} On the left we have theory corresponding to compactification on sphere with two maximal punctures with $SU(3)$ symmetry and two punctures with no symmetry. The former denoted by circles and latter by a cross and referred to as empty punctures. Dashed lines correspond to bidundamental chiral fields for which a baryonic superpotential is turned on. The squares correspond to $SU(3)$ global symmetry. On the right there is a trifundamental. It does not correspond to a compactification by itself, however gluing it to a theory by gauging $SU(3)$ corresponds to removing an empty puncture and adding a maximal puncture. From these blocks any theory in the class discussed here can be constructed. Note that both models here are equivalent in the IR but it is important to distinguish them to write a precise dictionary between compactifications and four dimensional models.}}

\

Using these blocks we can construct theories corresponding to any surface.  The construction proceeds iteratively by building bigger theories by gluing two smaller ones at a maximal puncture. The gluing is done by gauging with ${\cal N}=1$  vector multiplet the diagonal $SU(3)$ symmetry associated to the punctures. In Figure three two important examples are depicted.  These are related by RG flow we have discussed. In the geometric picture the vacuum expectation value we consider removes a maximal puncture and exchanges it with empty puncture.  We note that gluing theories corresponding to general surfaces all the $U(1)$ symmetries are broken, either by the superpotentials, anomalies, or both. In particular we are in general in the setup discussed above where we have theories obtained by gluing smaller pieces with a trifundamental so it is not charged under any $U(1)$.
 The theories have large conformal manifolds on which all the symmetries are broken. Theories with maximal punctures reside on same conformal manifolds as theories with only empty punctures such that every maximal puncture is traded with three empty punctures. For example, the theory with four maximal punctures is the same as one with twelve empty ones.  A general theory is then built from trifundamentals and baryonic superpotentials.  The superconformal R symmetry of all the chiral fileds is the free one.

\
 
 \centerline{\figscale{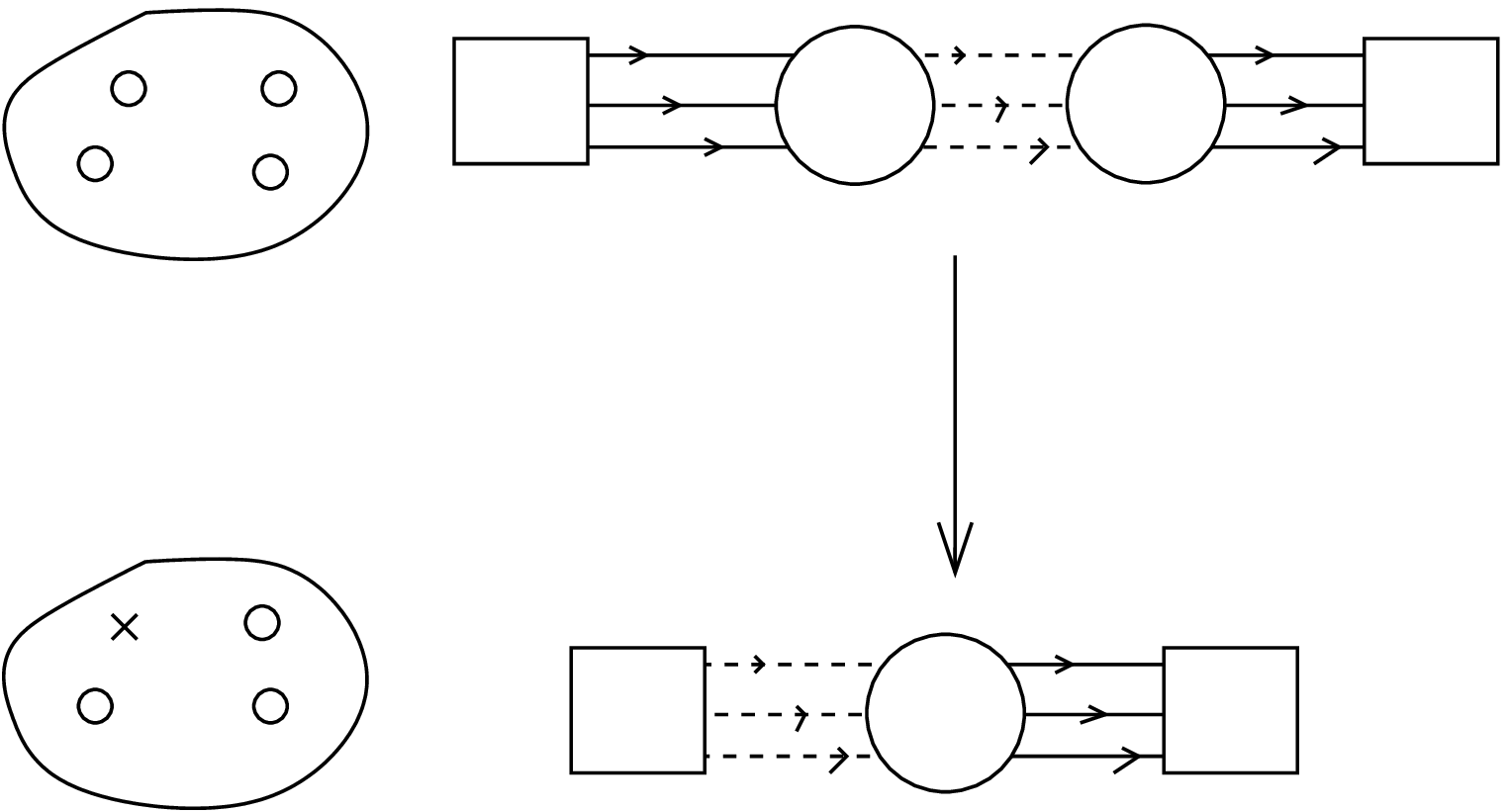}{2.8in}}
\medskip\centerline{\vbox{
\baselineskip12pt\advance\hsize by -1truein
\noindent\footnotefont{\bf Fig.~3:} On the top we combine a theory with two maximal punctures and two empty ones with two trifundamentals resulting in a sphere with four maximal punctures from which any surface with even number of punctures can be constructed. The fact that we can get only even number of punctures is as the punctures carry a $\Z_2$ valued twist.
 On the bottom we give vacuum expectation value to one of the baryons built from a field with no superpotential and obtain a theory with three maximal and one empty puncture.}}

With this we are ready to study the supersymmetric index and in particular the defects.

\subsec{The defect and the index}

Turning on a vacuum expectation value to an operator implies that we break the symmetry of the model in such a way that in IR the operator has zero charges. In the computation of the superconformal index \KinneyEJ\ this implies that we define fugacities in such a way that the weight of the operator in the computation is one. Doing so produces a pole divergence of the index and the claim \GaiottoXA\ is that the index of the theory in the IR is the residue of the pole. 

In what follows we will compute such residues for the relevant poles. We use standard index notations detailed in \RastelliTBZ\ with the different relevant functions defined in the Appendix.

 Let us denote index of some theory by ${\cal I}(y)$ with $y$ standing for $(y_1,y_2,y_3)$ such that $\prod_{i=1}^3 y_i=1$ being the fugacities for the maximal torus of $SU(3)$ flavor symmetry of one of the maximal punctures. The index depends on fugacities for all the symmetries associated to punctures but we leave those implicit in the definition. The index of a theory obtained by gluing a trifundamental field is by usual rules of index computations given by, 

\eqn\lsjfg{
{\cal I}' = (q;q)^2(p;p)^2\frac16\oint\frac{dy^1}{2\pi i y^1}\oint\frac{dy^2}{2\pi i y^2} \frac{\prod_{i,j,l=1}^3\Gamma_e((qp)^{\frac13} b_i y^j z_l^{-1})}{\prod_{i<j}\Gamma_e((y^i/y^j)^{\pm1})} {\cal I}(y)\,.
} The integration contours are around unit circle and we assume that parameters satisfy $|q|, |p|<1$ and $|b_i|=|z_i|=1$. We will soon take the latter to have more general values but then the contours should be properly deformed. Here the denominator comes from the vectors and the numerator from the trifundamental  field. The fugacities $b$ and $z$ are for the two $SU(3)$ symmetries of the trifundamental chiral field which become global symmetries of the new theory.  We next want to close one of these  two maximal punctures with $SU(3)$ symmetries by giving an expectation value to a baryon. By general considerations we have detailed the resulting theory should be the same as the model we started with plus a defect if the vacuum expectation value has a non trivial profile.  The vacuum expectation value corresponds to a pole in the index in the $SU(3)$ fugacities and we will study poles in $b_i$. The poles in $b_i$ of ${\cal I}'$ occur as when we vary the value of $b_i$ the poles in the integrated variables $z$ can pinch the integration contour \GaiottoXA\ (see also \AldayKDA\GaiottoUSA\MaruyoshiCAF\NazzalBRC\ItoFPL\ for similar computations). We turn to the  analysis of  such pinchings. 

The integrand has poles in $y^1$ such that the following are inside the unit circle,

\eqn\dlkjth{
y^1=(qp)^{\frac13} (y^2)^{-1} z_{j_3}^{-1}b_{k_3} q^{l_3} p^{n_3}\,,
} and the following are outside of the unit circle,

\eqn\dkjthg{
y^1=(qp)^{-\frac13} z_{j_1}b_{k_1}^{-1} q^{-l_1} p^{-n_1}\,.
} Similarly we have poles in $y^2$ outside the unit contour,

\eqn\krhg{
y^2=(qp)^{-\frac13} z_{j_2} b_{k_2}^{-1}q^{-l_2}p^{-n_2}\,.
} The numbers $l_i\,,\; n_j\,, \; k_l$ are non--negative integers.
From here if we want both integration contours to be pinched we have to satisfy in particular,

\eqn\ifugh{
1=q^{1+\sum_{h=1}^3 l_h}  p^{1+\sum_{y=1}^3 n_y} \prod_{r=1}^3 z_{j_r}^{-1}\prod_{c=1}^3 b_{k_c}\,.
}  This is the weight of a derivative of a baryon. Setting it to $1$ thus gives it vacuum expectation value.
We want to consider general class of poles which are independent of $z$ and then the indices  $j_r$ need to run over the three different values. Without loss we can take $z_{j_r}$ as $z_r$. When we will evaluate the residue we will need to sum over all the choices give by permutations of the three $z_r$, which will contribute an overall factor in the computation. We will ignore overall factors which do not depend on flavor fugacities in what follows as they will not be essential to the claims.
To obtain an interesting pole let us consider taking all $b_{k_i}$ to be the same, say $b_1$. Then we obtain that we have pinchings when,

\eqn\kjfghke{
b_1=q^{-\frac13(1+\sum_{h=1}^3 l_h)}p^{-\frac13(1+\sum_{y=1}^3 n_y)}\,.
} This corresponds to vacuum expectation value to $\det Q_1$. 
The poles are thus classified by two integers $\sum_{h=1}^3 l_h=L$ and $\sum_{y=1}^3 n_y=M$. These correspond to giving vacuum expectation values of the form $\partial_1^L\partial_2^M \det Q_1$. The two derivatives are the rotations  in  two orthogonal complex planes. The physical interpretation of the IR fixed point is of original theory, index of which is ${\cal I}(y)$, with two
 surface defects, each wrapping one of the equators of $\S^3$ and the $\S^1$.\foot{We parametrize the sphere as $\sum_{j=1}^4 x_j^2=1$. The two rotations are in the planes $(x_1 ,\, x_2)$ and $(x_3 ,\, x_4)$. The defects wrap either $x_1=x_2=0$ or $x_3=x_4=0$.}   Here we refer to the fact that the index can be thought as supersymmetric partition function on $\S^3\times \S^1$. One of the defects is determined by $L$ and another by $M$, see \GaiottoXA.

It is straightforward to compute the residues for general $L$ and $M$. Here we will do so in two simplest cases.
Let us take $L$ and $M$ to be zero. This corresponds to a vacuum expectation value for the baryon. This implies,

\eqn\jhgfsjk{
 y^i = z_i\,, \;\;\;\;  
} The integrand becomes, 

\eqn\efkjhgek{
\frac{\prod_{i,j}\Gamma_e(z_i/z_j)}{\prod_{j\neq i} \Gamma_e(z_i/z_j)} \prod_{i,j} \Gamma_e((q p)^{\frac13} b_2 z_i/z_j)\Gamma_e((q p)^{\frac23} (b_2)^{-1} z_i/z_j) {\cal I}(z_i)\propto \Gamma_e(1)^3 {\cal I}(y)\,.
}
Note that $\Gamma_e(z\to 1) \to \infty$ as $1/(1-z)$. We have thus third order singularity which can be thought as three simple poles colliding. Two of them are absorbed by the two contour integrals and the third one is the pole in $b$ we are after.
We have,

\eqn\rljghk{
Res_{b_1\to (q p)^{-\frac13}} {\cal I}'\propto {\cal I}(z_i)\,.
} Note that although we only broke explicitly $b_1$ all fields charged under $b_2$ acquired mass and decoupled in IR. This happens as in \efkjhgek\  we observe that,

\eqn\slkghlwrglwrt{
\Gamma_e((q p)^{\frac13} b_2 z_i/z_j)\Gamma_e((q p)^{\frac23} (b_2)^{-1} z_i/z_j)=1\,,} with the two $\Gamma_e$ corresponding to the index of $Q_2$ and $Q_3$ which is consistent with having a mass term.
This will change when we turn on vacuum expectation value for derivatives of the operator.

Let us consider then  the case with $L=1$ and $M=0$. We can satisfy this by taking one of the $l_i=1$ and rest zero. Different choices will differ by permutations of $z_i$ and we will need to sum over all of them eventually. Let us take $l_1=1$ and all others zero. We thus have,

\eqn\kdjhgis{
b_1=(q p)^{-\frac13}q^{-\frac13}\,,\;\;\; y^1 = z_1q^{-\frac23}\,,\;\;\;\; y^2 =z_2q^{\frac13}\,,\;\;\;\; y^3 =z_3 q^{\frac13}\, .
} We have then for the trifundamental matter,

\eqn\kdjfhgke{\eqalign{
&\Gamma_e(q^{-1}z_1/z_i)\Gamma_e(z_2/z_i)\Gamma_e(z_3/z_i)\Gamma_e((q p)^{\frac13}q^{-\frac23}b_2 z_1/z_i)\
\Gamma_e((q p)^{\frac13}q^{\frac13}b_2 z_2/z_i)\cr &
\Gamma_e((q p)^{\frac13}q^{\frac13}b_2 z_3/z_i) \Gamma_e((q p)^{\frac23}q^{-\frac13}b_2^{-1} z_1/z_i)
\Gamma_e((q p)^{\frac23}q^{\frac23}b_2^{-1} z_2/z_i)\Gamma_e((q p)^{\frac23}q^{\frac23}b_2^{-1}   z_3/z_i)\,.
}
}  Introducing the contribution of the vector and ${\cal I}(y)$ the residue evaluates, ignoring as usual overall factors, to ,

\eqn\ksjrghklas{
\frac{\theta_p((p/q)^{\frac13} b_2 z_2/z_3)\theta_p((p/q)^{\frac13} b_2 z_3/z_2)}{\theta_p(z_2/z_1)\theta_p(z_3/z_1)} {\cal I}(z_1 q^{-\frac23},z_2 q^{\frac13}, z_3 q^{\frac13})\,.
}
All in all summing over different choices of $l_i$ we have that,

\eqn\ewlfhgk{
Res_{b_1\to (q p)^{-\frac13} q^{-\frac13}} {\cal I}' \propto {\cal O}^{b_2 (q p)^{-\frac16} q^{-\frac16}}_q(z) \cdot {\cal I}(z)\,.
} Here we define,

\eqn\ekjrhgk{\eqalign{
&{\cal O}^{Y}_q(z)\cdot {\cal I}(z)  = \frac{\theta_p(p^{\frac12}  Y (z_2/z_3)^{\pm1})}{\theta_p( z_2/z_1)\theta_p(z_3/z_1)} {\cal I}(z_1 q^{-\frac23}, z_2  q^{\frac13},z_3 q^{\frac13})+\cr
&\;\;\;\;\frac{\theta_p(p^{\frac12}  Y (z_1/z_3)^{\pm1})}{\theta_p( z_1/z_2)\theta_p(z_3/z_2)} {\cal I}(z_1 q^{\frac13}, z_2  q^{-\frac23},z_3 q^{\frac13})+\frac{\theta_p(p^{\frac12}  Y (z_2/z_1)^{\pm1})}{\theta_p( z_2/z_3)\theta_p(z_1/z_3)} {\cal I}(z_1 q^{\frac13}, z_2  q^{\frac13},z_3 q^{-\frac23})\,.
}
} This is the difference operator introducing a surface defect into the index computations.
Note that the operator now depends on the $SU(2)_f$ commutant of $b_1$ in $SU(3)$ fugacity for which is $Y$. When we turn on vacuum expectation values for a derivative of a baryon operator some of the states localized on the defect are charged under this $SU(2)_f$ symmetry though there are no bulk states charged under the symmetry. The expression \ewlfhgk\ gives the index of the theory in the IR in presence of the defect and thus it explicitly shows that the defect carries the $SU(2)_f$ flavor symmetry. Each defect will have a factor of $SU(2)_f$ associated to it.

Note that the theta functions appearing in the numerator in the difference operator above have natural interpretation as elliptic genus of ${\cal N}=(2,0)$ Fermi multiplet and the  theta functions in denominator as the elliptic genus of a chiral field, see \BeniniNDA. It will be interesting to find a two dimensional CFT which when coupled to the four dimensional model is equivalent to our defect. We leave this for future work.

We can repeat the procedure with other values of $M$ and $L$ and derive operators corresponding to other surface defects. Let us just mention that the operator corresponing to $M=1$ and $L$ being zero is the same as above with $q$ and $p$ exchanged, and we will denote it by ${\cal O}^Y_p$. The two operators introduce same type of defects but on different two dimensional locus.

\

\subsec{ Checks of the compactifications using the operator}

Using this result we can  subject the map between the compactifications and four dimensional models suggested in \RazamatGRO\ to numerous checks. As the index is independent on the couplings, it needs to be the same in all duality frames. In particular if the theory corresponds to compactification with several maximal punctures, acting with the difference operator on fugacities of different symmetries, and in any order, should produce the same result. The latter implies that all the operators have to commute.
We claim that,

\eqn\ldfjkhgke{
[{\cal O}^{X}_q\,,\;\; {\cal O}^{Z}_q]=0\,.
} This is a non trivial fact.
This will follow from the following theta function identity,

\eqn\ekrjghkw{	\eqalign{&
\frac{\theta_p(\frac{z q z_1}{z_2}) \theta_p(\frac{z z_2}{q z_1}) \theta_p(\frac{x z_2}{z_3}) \theta_p(\frac{x z_3}{z_2})}{\theta_p(\frac{z_1}{z_3})\theta_p(\frac{z_3}{q z_1})}+\frac{\theta_p(\frac{x z_1}{z_2})  
\theta_p(\frac{x z_2}{z_1})\theta_p(\frac{z z_2}{q z_3})\theta_p(\frac{z q z_3}{z_2})}{\theta_p(\frac{z_1}{q z_3}) \theta_p(\frac{z_3}{z_1})}=\cr &
\frac{\theta_p(\frac{x q z_1}{z_2}) \theta_p(\frac{x z_2}{q z_1}) \theta_p(\frac{z z_2}{z_3}) \theta_p(\frac{z z_3}{z_2})}{\theta_p(\frac{z_1}{z_3})\theta_p(\frac{z_3}{q z_1})}+\frac{\theta_p(\frac{z z_1}{z_2}) \theta_p(\frac{z z_2}{z_1})\theta_p(\frac{x z_2}{q z_3})\theta_p(\frac{x q z_3}{z_2})}{\theta_p(\frac{z_1}{q z_3}) \theta_p(\frac{z_3}{z_1})}\,.}
}  All the operators, for different values of $M$ and $L$,  should commute and in particular,

\eqn\lekfuh{
[{\cal O}^X_q\,,\;\; {\cal O}^Z_p]=0\,.
} This follows from elementary identities, for example $\theta_p(p z)=\theta_p(1/z)$ and $\theta_p (1/z)=-1/z\theta_p(z)$. We can act with the operators on our basic building block, the trifundamental, and it should not matter on which fugacity we apply the operator. In mathematical jargon the index of the trifundamental is a {\it kernel function}, see for example \noumi, 
 of the difference operators,

\eqn\elrjhgl{\eqalign{
&
{\cal O}^Y_p(z)\cdot \prod_{i,j,l=1}^3\Gamma_e((q p)^{\frac13} z_i^{-1} y_j^{-1} b_l^{-1}) =\cr
&
 {\cal O}^Y_p(y)\cdot \prod_{i,j,l=1}^3\Gamma_e((q p)^{\frac13} z_i^{-1} y_j^{-1} b_l^{-1}) = \cr
 &
 {\cal O}^Y_p(b)\cdot \prod_{i,j,l=1}^3\Gamma_e((q p)^{\frac13} z_i^{-1} y_j^{-1} b_l^{-1}) \,.
 }
} Note that $\prod_i z_i =\prod_j y_j=\prod_l b_l=1$. This equality reduces to an identity of sum of products of theta functions. We have not proven this but checked it using {\it Mathematica} in expansion in the  fugacities $p$ and $q$. 

The operator is self-adjoint under the vector multiplet measure,

\eqn\wufhkw{\eqalign{&
\oint \frac{dz_1}{2\pi i z_1}\oint \frac{dz_2}{2\pi i z_2}\prod_{i<j}\frac1{\Gamma_e((z_i/z_j)^{\pm1})} f(z^{-1})\left[ {\cal O}^Y_q(z) \cdot h(z)\right]=\cr&
\oint \frac{dz_1}{2\pi i z_1}\oint \frac{dz_2}{2\pi i z_2}\prod_{i<j}\frac1{\Gamma_e((z_i/z_j)^{\pm1})}  \left[{\cal O}^Y_q(z^{-1})\cdot f(z^{-1})\right]  h(z)\,.
}
} This implies that if we prove the kernel function property the action of the operator will be independent of the choice of the maximal puncture for any theory as we will be able to pull the operator through the integrals.
The self-adjointness can be easily shown with the assumptions that the functions do not have poles in some strip around the unit circle.  For example, let us look at one of the three terms in the integrand,

\eqn\lskfghkw{
\oint \frac{dz_1}{2\pi i z_1}\oint \frac{dz_2}{2\pi i z_2 }\prod_{i<j}\frac1{\Gamma_e((z_i/z_j)^{\pm1})} f(z^{-1})  \frac{\theta_p(p^{\frac12}  Y (z_2/z_3)^{\pm1})}{\theta_p( z_2/z_1)\theta_p(z_3/z_1)}g(z_1 q^{-\frac23}, z_2  q^{\frac13},z_3 q^{\frac13})\,.
} We perform change of variables with $z_1\to z_1 q^{\frac23}$ and  $z_2 \to z_2  q^{-\frac13}$. Then the term becomes,

\eqn\lrjghkrwle{\eqalign{
&\oint \frac{dz_1}{2\pi i z_1}\oint \frac{dz_2}{2\pi i z_2 }\frac1{\Gamma_e((z_2/z_3)^{\pm1})}\frac1{\Gamma_e(q z_1/z_2)\Gamma_e(q z_1/z_3)}\frac1{\Gamma_e(q^{-1} z_2/z_1)\Gamma_e(q^{-1} z_3/z_1)}\cr&
 f(z_1^{-1}  q^{-\frac23}, z_2^{-1}  q^{\frac13},z_3^{-1} q^{\frac13})  \frac{\theta_p(p^{\frac12}  Y (z_2/z_3)^{\pm1})}{\theta_p( q^{-1} z_2/z_1)\theta_p(q^{-1} z_3/z_1)}g(z)\,.
 }
 } We now use that $\Gamma_e(q z)=\theta_p(z) \Gamma_e(z)$ to write the above,
 
 \eqn\kwrjghkw{
 \;\;\;\oint \frac{dz_1}{2\pi i z_1}\oint \frac{dz_2}{2\pi i z_2 }\prod_{i<j}\frac1{\Gamma_e((z_i/z_j)^{\pm1})} f(z_1^{-1}  q^{-\frac23}, z_2^{-1}  q^{\frac13},z_3^{-1} q^{\frac13}) \frac{\theta_p(p^{\frac12}  Y (z_2/z_3)^{\pm1})}{\theta_p(  z_1/z_2)\theta_p(z_1/z_3)}g(z)\,.
} This shows the self-adjointness. 

We thus find that the properties of the defects are consistent with the conjectured map between compactifications and four dimensional theories of \RazamatGRO\ and this  is a non trivial check of that suggestion.

\

\newsec{Case of $SO(8)$ minimal conformal matter} 

We repeat the analysis of the previous section for the (twisted) compactifications of the minimal conformal matter of type $SO(8)$. The analysis is very similar but the details are a bit different. We refer again to \RazamatGRO\ for details.
First the general set of models is obtained by gluing together theories with $SU(4)$ gauge groups. The basic building block is  a {\it pair} of bifundamental fields of $SU(4)$ which we will denote by $\widetilde Q_i$. The pair $\widetilde Q_i$ is rotated by $SU(2)_f$, has R charge half, and is not charged under any $U(1)$ symmetry.
  Here we consider turning on vacuum expectation values to $\det \widetilde Q_1$ and derivatives of this. The  $SU(2)_f$ is broken by the vacuum expectation value. The chiral field $\widetilde Q_2$ in the IR acquires a mass and decouples.
   Thus in this case we will obtain surface defects which do not carry any flavor symmetry, see  Figure four for illustration.
   
\

\centerline{\figscale{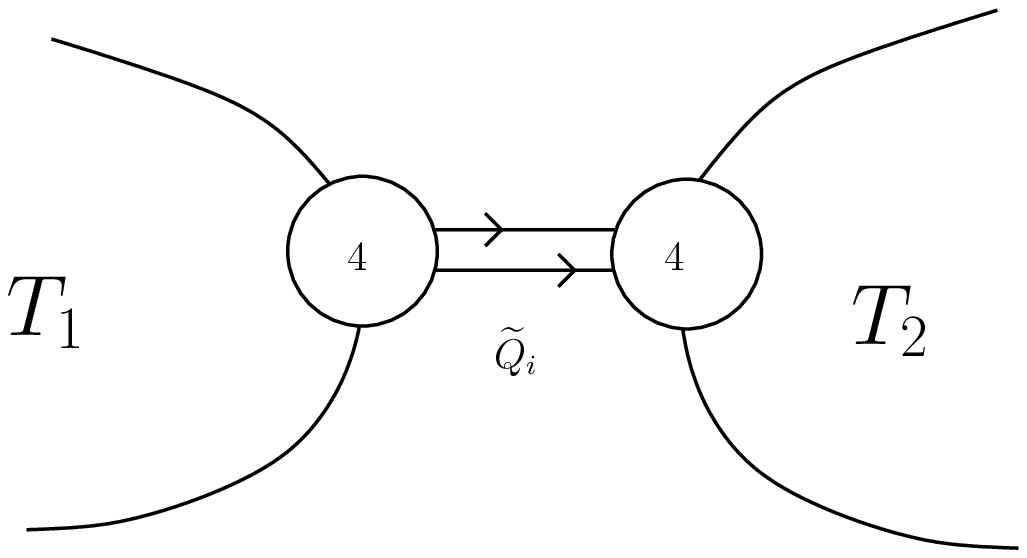}{2.4in} $\;\;\;\;\;\qquad \qquad $  \figscale{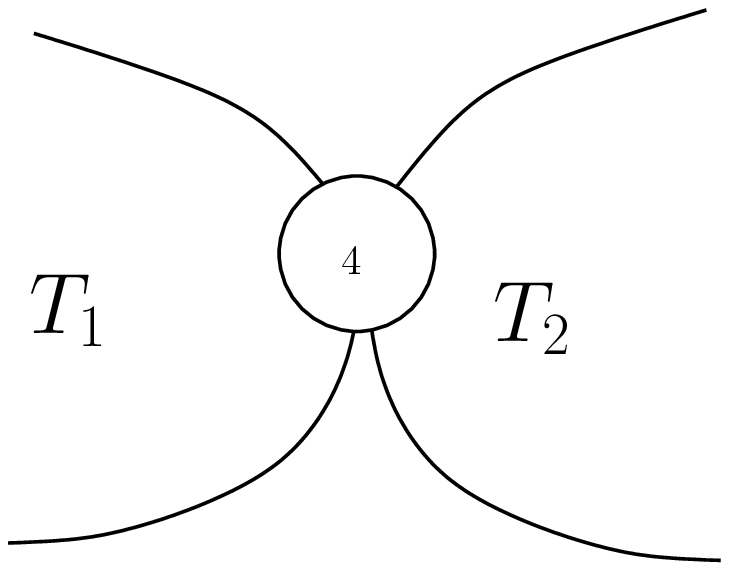}{1.6in}}
\medskip\centerline{\vbox{
\baselineskip12pt\advance\hsize by -1truein
\noindent\footnotefont{\bf Fig.~4:} Two theories $T_1$ and $T_2$ combined by gauging two $SU(4)$ symmetries connected by two bifundamental chiral fields $\widetilde Q_i$.  We assume that there is no $U(1)$ symmetry under which $\widetilde Q_i$ are charged.
We give vacuum expectation value to a baryon $\det \widetilde Q_1$. The theory in the IR is the two theories glued by gauging diagonal combination of two $SU(4)$ symmetries. The vacuum expectation value breaks $SU(2)_f$ symmetry rotating the two $Q_i$.}}

 \

Before turning to the derivation of the quantum mechanical model introducing the defects to the index computation we review the basic properties of the map between the compactifcation of the minimal conformal matter of type $SO(8)$ and the four dimensional gauge theories derived in \RazamatGRO. Here the basic building block is $\widetilde Q_i$ with a baryonic superpotential preserving the two $SU(4)$ symmetries. It corresponds to a sphere with two maximal punctures carrying $SU(4)$ symmetry, and two basic punctures which carry no symmetry. The bifundamental with no baryonic superpotential does not correspond by itself to a compactification but when glued to another theory by gauging one of the $SU(4)$ symmetries it adds a new type of puncture carrying $SU(2)$ symmetry,\foot{For some curiosities in defining punctures in this set of models see \RazamatGRO.} which we will denote as $\widehat{SU}(2)$ puncture, and removes one basic puncture. See Figure five for illustration. As in the previous case the generic models have large conformal manifolds on which all the symmetries are broken. The statement is that theories with a maximal punctures reside on the same conformal manifold as the theory with the puncture traded by two $\widehat{SU}(2)$ punctures or six basic punctures. The $\widehat{SU}(2)$ punctures can be traded by three basic punctures. 

\

 \centerline{\figscale{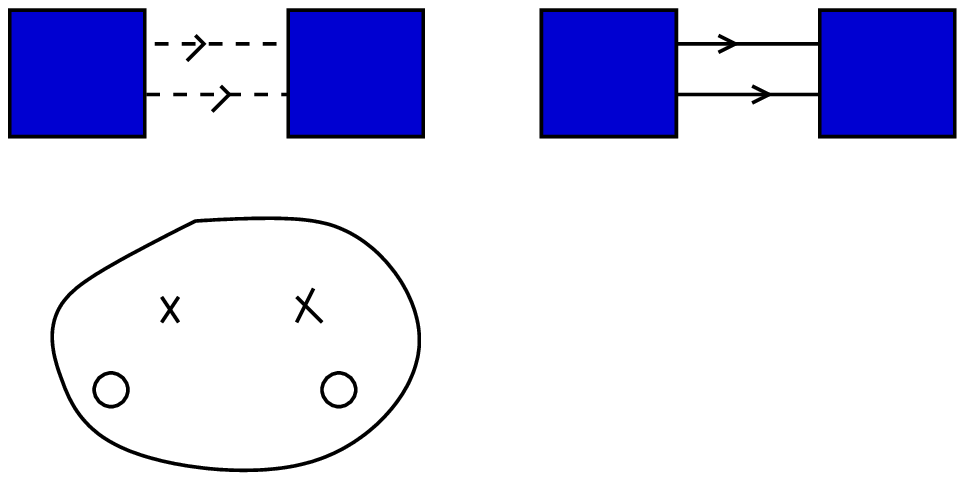}{2.4in}}
\medskip\centerline{\vbox{
\baselineskip12pt\advance\hsize by -1truein
\noindent\footnotefont{\bf Fig.~5:} On the left we have theory corresponding to compactification on sphere with two maximal punctures with $SU(4)$ symmetry and two punctures with no symmetry. The former denoted by circles and latter by a cross and referred to as basic punctures. Dashed lines correspond to bidundamental chiral fields for which a baryonic superpotential is turned on. The colored squares correspond to $SU(4)$ global symmetry. On the right there is a pair of bifundamentals. It does not correspond to a compactification by itself, however gluing it to a theory by gauging $SU(4)$ corresponds to removing a basic puncture and adding an $\widehat{SU}(2)$ puncture. From these blocks any theory can be constructed.}} 

\

In this class of theories we can construct a three punctured sphere with three maximal punctures by combining two bifundamentals with no baryon superpotential and one with the superpotential, see Figure six. Note that here naively the model has $SU(4)\times SU(4)\times SU(2)\times SU(2)$ symmetry but the conjecture of \RazamatGRO\ is that somewhere on the conformal manifold the symmetry enhances to $SU(4)\times SU(4)\times SU(4)$. From this block theories corresponding to any Riemann surface can be constructed. The flows we consider giving vacuum expectation values to the baryons close $\widehat{SU}(2)$ puncture to the basic puncture.

We now are ready to discuss the defects in the index computation.

\

\subsec{The defect and the index} 

In this case we denote the index of general theory with a maximal puncture by ${\cal I}(y)$ with the $y_i$ being fugacities for $SU(4)$. We glue to it $\widetilde Q_i$ which has an additional $SU(4)$ parametrized by $z_i$ and an $SU(2)$ parametrized by $b$. The index of the combined model is given by
the following integral,

\eqn\lsjfg{
{\cal I}' = (q;q)^3(p;p)^3\frac1{24}\oint\frac{dy^1}{2\pi i y^1}\oint\frac{dy^2}{2\pi i y^2}\oint\frac{dy^3}{2\pi i y^3} \frac{\prod_{l,j =1}^4\Gamma_e((qp)^{\frac14} b^{\pm1} y^j z_l^{-1})}{\prod_{i<j}\Gamma_e((y^i/y^j)^{\pm1})} {\cal I}(y)\,.
} We are interested in poles of the index in $b$, which is in closing the $SU(2)$ puncture. We will perform the analysis of the divergences in $b$ and discuss interesting set of poles corresponding to vacuum expectation values to derivatives of baryon $\det \widetilde Q_1$. 
The integrand has poles in $y^1$ such that the following are inside the unit circle,

\

 \centerline{\figscale{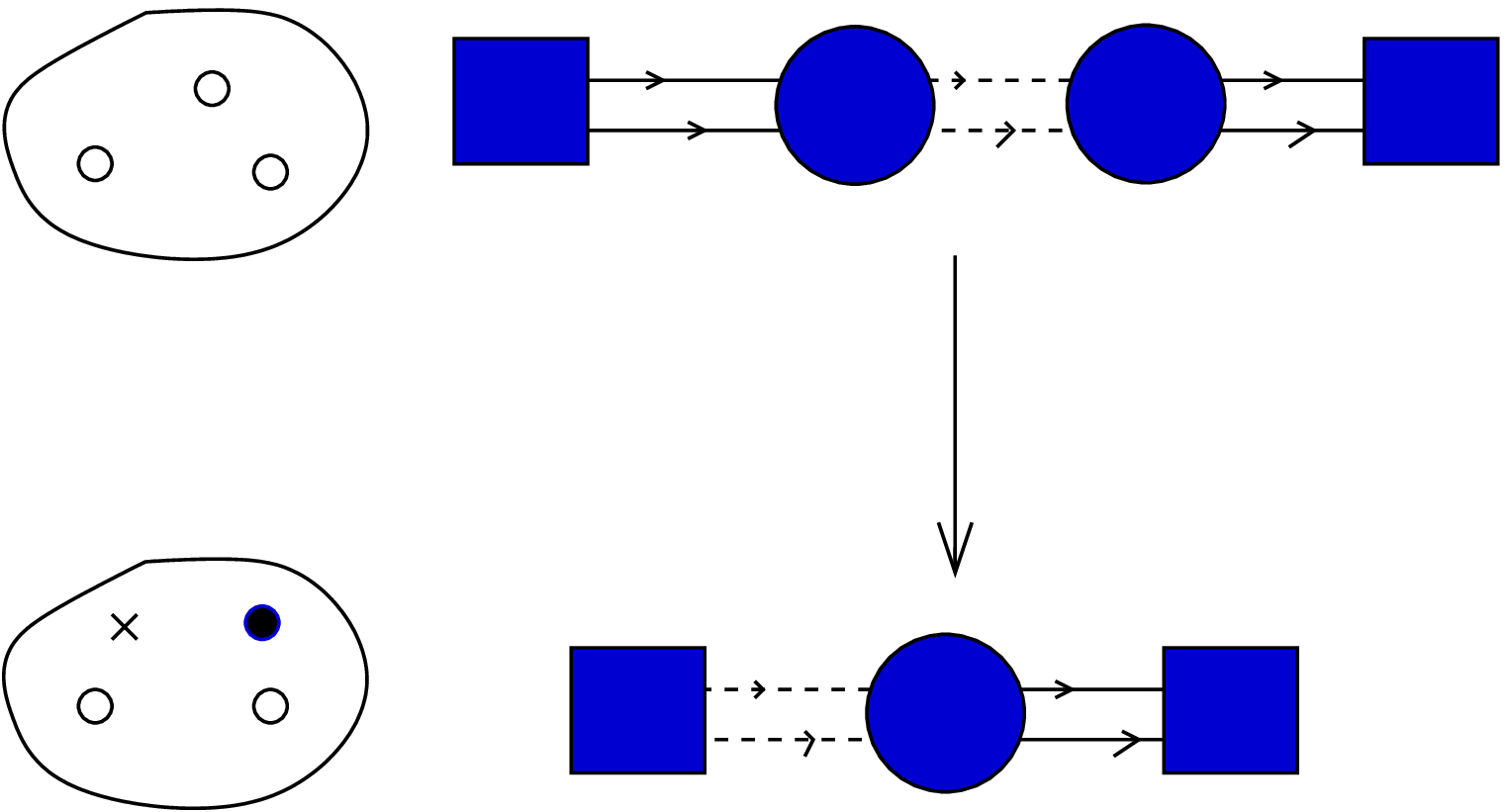}{2.6in}}
\medskip\centerline{\vbox{
\baselineskip12pt\advance\hsize by -1truein
\noindent\footnotefont{\bf Fig.~6:} On the top we combine a theory with two maximal punctures and two basic ones with two ${} \; \widetilde Q_i$ resulting in a sphere with three maximal punctures from which any surface  can be constructed. On the bottom we give vacuum expectation value to one of the baryons built from a field with no superpotential and obtain a theory with two maximal, one basic puncture, and one $\widehat{SU}(2)$ puncture. We denote the $\widehat{SU}(2)$ puncture with colored circle.}}

\

\eqn\dlkjth{
y^1=(qp)^{\frac14} (y^2)^{-1} (y^3)^{-1} z_{j_4}^{-1} b^{\pm1} q^{l_4} p^{n_4}\,,
} and the following are outside of the unit circle,

\eqn\dkjthg{
y^1=(qp)^{-\frac14} z_{j_1}b^{\pm1} q^{-l_1} p^{-n_1}\,.
} Similarly we have poles in $y^2$ and $y^3$ outside the unit contour,

\eqn\krhg{\eqalign{
& y^2=(qp)^{-\frac14} z_{j_2} b^{\pm1}q^{-l_2}p^{-n_2}\,,  \cr
& y^3=(qp)^{-\frac14} z_{j_3} b^{\pm1}q^{-l_3}p^{-n_3}\,.
}
} From here a necessary condition if we want all the integration contours to pinch will be,

\eqn\ifugh{
1=q^{1+\sum_{h=1}^4 l_h}  p^{1+\sum_{y=1}^4 n_y} \prod_{r=1}^4 z_{j_r}^{-1}b^x\,.
}  This is the weight of a derivative of a baryon. Setting it to $1$ thus gives it vacuum expectation value.
As we want general poles independent of $z$  will take without loss $z_{j_r}=z_r$. We will also take $x=4$ which is corresponds to vacuum expectation values for derivatives of $\det \,Q_1$ (note that it could be $-4,-2,0,2,4$).
The we have a pole when,

\eqn\lrejhgk{
b=(q p)^{-\frac14}q^{-\frac{M}4}p^{-\frac{L}4}\,.
} First we analyze the case of both $L$ and $M$ $=0$. We have,

\eqn\elrighlw{
y^i=z_i\,.
} The integrand evaluates to,

\eqn\keurhglrg{
 \; \; \frac{\prod_{j,l=1}^4\Gamma_e((qp)^{\frac12}   z_j/z_l) \Gamma_e(  z_j/z_l)}{\prod_{i<j} \Gamma_e(z^i/z^j)  \Gamma_e(z^j/z^i)} {\cal I}(z) \to \Gamma_e(1)^4{\cal I}(z)\,. 
 }
 We see that we have order four divergence which gives us a pole and the residue is (as always ignoring overall factors),

\eqn\wrkjgh{
Res_{b\to (q p)^{-\frac14}} {\cal I}'\propto {\cal I}(z_i)\,.
} Now we take $M$ to be $1$ and $L$ to be $=0$. We have taking $l_1=1$,

\eqn\lekjhgek{
y^1=z_1q^{-\frac34}\,,\;\;\;\; y^2=z_2 q^{\frac14}\,, \;\;\;\; y^3=z_3 q^{\frac14}\,,\;\;\;  y^4=z_4 q^{\frac14} \,.
} The integrand becomes,

\eqn\lwkrhgwligjw{
\frac{\Gamma_e(q^{-1}  z_1/z_j)\Gamma_e(z_j/z_l) \Gamma_e((q p)^{\frac12} q^{-\frac12} z_1/z_j)\Gamma_e((q p)^{\frac12} q^{\frac12} z_j/z_l) }{\Gamma_e(q^{-1} z_1/ z_i)\Gamma_e(q z_i/z_1)\Gamma_e(z_i/z_j)} {\cal I}(z_1 q^{-\frac34}, z_2 q^{\frac14}, z_3 q^{\frac14},z_4 q^{\frac14})\,.
} Evaluating the residue we get removing overall factors,

\eqn\elrjthekr{
\frac{ \theta_p(p^{\frac12}  z_3/z_2)\theta_p(p^{\frac12}  z_4/z_2) \theta_p(p^{\frac12}  z_3/z_4)  }{\theta_p(z_2/z_1)\theta_p(z_3/z_1)\theta_p(z_4/z_1)} {\cal I}(z_1 q^{-\frac34}, z_2 q^{\frac14}, z_3 q^{\frac14},z_4  q^{\frac14})\,.
}  We obtain that,

\eqn\lejrhgeklrhg{
Res_{b\to ( q p)^{-\frac14}q^{-\frac14}} {\cal I}' \propto {\cal O}_q(z) \cdot {\cal I}(z)\,.
} We have,

\eqn\krjghkw{\eqalign{
&{\cal O}_q(z)\cdot {\cal I}(z) = \frac{ \theta_p(p^{\frac12}  z_3/z_2)\theta_p(p^{\frac12}  z_4/z_2) \theta_p(p^{\frac12}  z_3/z_4)  }{\theta_p(z_2/z_1)\theta_p(z_3/z_1)\theta_p(z_4/z_1)} {\cal I}(z_1 q^{-\frac34}, z_2 q^{\frac14}, z_3 q^{\frac14},z_4  q^{\frac14})+\cr
&\;\;\;\;\;\;\;\frac{ \theta_p(p^{\frac12}  z_3/z_1)\theta_p(p^{\frac12}  z_1/z_4) \theta_p(p^{\frac12}  z_3/z_4)  }{\theta_p(z_1/z_2)\theta_p(z_3/z_2)\theta_p(z_4/z_2)} {\cal I}(z_1 q^{\frac14}, z_2 q^{-\frac34}, z_3 q^{\frac14},z_4  q^{\frac14})+\cr
&\;\;\;\;\;\;\frac{ \theta_p(p^{\frac12}  z_1/z_2)\theta_p(p^{\frac12}  z_4/z_2) \theta_p(p^{\frac12}  z_1/z_4)  }{\theta_p(z_2/z_3)\theta_p(z_1/z_3)\theta_p(z_4/z_3)} {\cal I}(z_1 q^{\frac14}, z_2 q^{\frac14}, z_3 q^{-\frac34},z_4  q^{\frac14})+\cr
&\;\;\;\;\;\;\;\frac{ \theta_p(p^{\frac12}  z_3/z_2)\theta_p(p^{\frac12}  z_1/z_2) \theta_p(p^{\frac12}  z_3/z_1)  }{\theta_p(z_3/z_4)\theta_p(z_2/z_4)\theta_p(z_1/z_4)} {\cal I}(z_1 q^{\frac14}, z_2 q^{\frac14}, z_3 q^{\frac14},z_4  q^{-\frac34})
\,.
}
}This is the difference operator introducing the surface defect into the index computation.
We can define more general operators taking $M$ and $L$ arbitrary. For $M=1$ and $L$ equal zero we get the same operator with $q$ and $p$ exchanged.
The operators ${\cal O}_q$ and ${\cal O}_p$ commute.
The operator ${\cal O}_q$ (and all the others) should be self-adjoint under vector multiplet measure,

\eqn\wsljrghjk{\eqalign{&
\oint \frac{dz_1}{2\pi i z_1}\oint \frac{dz_2}{2\pi i z_2}\oint \frac{dz_3}{2\pi i z_3} \prod_{i<j}\frac1{\Gamma_e((z_i/z_j)^{\pm1})} f(z^{-1})\left[ {\cal O}_q(z) \cdot h(z)\right]=\cr&
\oint \frac{dz_1}{2\pi i z_1}\oint \frac{dz_2}{2\pi i z_2}\oint \frac{dz_3}{2\pi i z_3} \prod_{i<j}\frac1{\Gamma_e((z_i/z_j)^{\pm1})}  \left[{\cal O}_q(z^{-1})\cdot f(z^{-1})\right]  h(z)\,.
}
} The proof for the operator \krjghkw\ is elementary as in the previous case.
The index of the $\widetilde Q_i$ is a kernel function for the operator,

\eqn\lrsjkhgkwe{
{\cal O}_p(z)\cdot \prod_{i,j=1}^4\Gamma_e((q p)^{\frac14} z_i^{-1} b^{\pm1} y_j^{-1}) ={\cal O}_p(y)\cdot \prod_{i,j=1}^4\Gamma_e((q p)^{\frac14} z_i^{-1} y_j^{-1} b^{\pm1}) \,.
 } Note that $\prod_i z_i=\prod_j y_j=1$. As in the previous case we checked this equality in expansion in fugacities. These properties guarantee that the action of the difference operator is consistent with the conjectured map between compactifications and four dimensional theories.

 \
 
 \

\noindent{\it Acknowledgments: } We would like to thank Simon Ruijsenaars and Gabi Zafrir 
 for comments and relevant discussions. The research was  supported by the Israel Science Foundation under grant No. 1696/15 and by the I-CORE  Program of the Planning and Budgeting Committee.

\appendix{A}{Definitions of functions}

We define the theta functions and the q-Pochhammer symbol,

\eqn\lerjhg{
\theta_q(z) =\prod_{j=1}^\infty (1- z q^{j-1})(1-1/z q^j)\,,\qquad (z;q) = \prod_{j=1}^\infty (1 - z q^{j-1})\,.
} The elliptic Gamma function is,

\eqn\kejrgh{
\Gamma_e(z)=\prod_{i,j=1}^\infty \frac{1-\frac{q p}z q^{j-1}p^{i-1}}{1-z q^{i-1} p^{j-1}}\,.
} We omit the parameters $q$ and $p$ from the definitions of the elliptic Gamma function for brevity. We also use the short-hand notations,

\eqn\ekwjfhk{
f(a z^{\pm1}) = f(a z) f(a z^{-1})\,.
} The theta functions and the elliptic Gamma functions satisfy many identities and here are some of them,

\eqn\ekrhgk{\eqalign{
&  \theta_p(p z) =\theta_p(1/z) = -\frac1z \theta_p(z)\,, \;\;\;\;\qquad \Gamma_e(qz)=\theta_p(z)\Gamma_e(z)\,,\cr
& \Gamma_e(qp/z)\Gamma_e(z)=1\,.
}
} See \Spiridonov\ for a useful reference on elliptic Gamma functions and their properties. 

\listrefs
\end